\documentclass[aps,prl,reprint,superscriptaddress,nofootinbib, longbibliography]{revtex4-2}

\usepackage[T1]{fontenc}
\usepackage{amsmath,amssymb,bm}
\usepackage{graphicx}
\usepackage{xcolor}
\usepackage{overpic}
\usepackage{caption}
\usepackage{tikz}
\usepackage{mathtools,booktabs}
\DeclarePairedDelimiter{\norm}{\lVert}{\rVert}

\usepackage[final]{microtype}

\makeatletter
\g@addto@macro\normalsize{%
  \setlength{\abovedisplayskip}{5pt plus 2pt minus 2pt}%
  \setlength{\belowdisplayskip}{5pt plus 2pt minus 2pt}%
  \setlength{\abovedisplayshortskip}{3pt plus 2pt minus 1pt}%
  \setlength{\belowdisplayshortskip}{3pt plus 2pt minus 1pt}%
}
\makeatother

\usepackage[
    colorlinks=true,
    linkcolor=red,
    citecolor=blue,
    urlcolor=blue
]{hyperref}

\makeatletter
\AtBeginDocument{%
  \long\def\@makecaption#1#2{%
    \par
    \vskip\abovecaptionskip
    \begingroup
      \small\rmfamily
      \linespread{0.95}\selectfont
      \sbox\@tempboxa{%
        \let\\\heading@cr
        \@make@capt@title{\textbf{#1}}{#2}%
      }%
      \ifdim\wd\@tempboxa>\hsize
        \begingroup
          \samepage
          \flushing
          \let\footnote\@footnotemark@gobble
          \@make@capt@title{\textbf{#1}}{#2}\par
        \endgroup
      \else
        \global\@minipagefalse
        \hb@xt@\hsize{\unhbox\@tempboxa\hfil}%
      \fi
    \endgroup
    \vskip\belowcaptionskip
  }%
}
\makeatother

\newcommand{\Mc}{\mathcal{M}_c}

\newcommand{\Like}{\mathcal{L}}

\newcommand{\flow}{f_{\rm low}}

\newcommand{\Msun}{M_\odot}

\newcommand{\tf}{t_f}

\begin{document}

\title{Fast Bayesian Inference for Long-Duration Gravitational-Wave Signals in 3G detectors}

\author{Anand S. Sengupta}
\affiliation{Indian Institute of Technology Gandhinagar, Gandhinagar 382055, India}
\email{asengupta@iitgn.ac.in}

\date{\today}

\begin{abstract}
Third-generation~(3G) gravitational-wave detectors will observe binary-neutron-star inspirals for nearly a day, so Earth’s rotation becomes part of the signal rather than a negligible correction. This destroys the usual separation between intrinsic and extrinsic parameters and creates a major computational bottleneck for Bayesian inference. We show that 
the exact five-harmonic Jaranowski--Kr\'olak--Schutz decomposition 
restores the separation of the rotating antenna-amplitude
response from the expensive intrinsic waveform calculations.
By sampling the intrinsic waveform on a frequency grid set by its phase curvature and then applying error-controlled relative binning, likelihood evaluations for a 21.3-h signal are accelerated by $\mathcal{O}(10^4)$, making day-long 3G signal inference \textcolor{black}{practical for the signal model considered here}.

\end{abstract}

\maketitle

\emph{\textbf{Introduction.}---}
Third-generation ground-based gravitational-wave detectors like Cosmic Explorer (CE)~\cite{evans2021horizonstudycosmicexplorer, Key_2026} and Einstein Telescope (ET)~\cite{Maggiore:2019uih} will observe compact-binary inspirals for durations far longer than those encountered in the current network of second-generation Advanced LIGO~\cite{Aasi:2014mqd}, Advanced Virgo~\cite{Acernese:2014hva}, and KAGRA~\cite{Akutsu:2020his} detectors.
For binary neutron stars, lowering the low-frequency seismic-wall from tens of
hertz to a few hertz extends the observable signal from minutes to tens of hours: a GW170817-like binary neutron star source enters the band at $\flow=2$~Hz roughly 21~hours (nearly a sidereal day) before merger.
Over such durations the detector does not remain fixed with respect to
the source. The Earth rotates appreciably, and the antenna response becomes a genuine time-dependent part of the signal model.

For the nonprecessing, dominant-$(2,2)$ signals considered here, the two polarizations are proportional to a common intrinsic waveform,
$\tilde h_{22}(f;\bm{\lambda})$, where $\bm{\lambda}$ denotes the
intrinsic parameters.  
We write
\begin{equation}
    \tilde h(f)
    =
    {\cal F} \:
        \tilde h_{22}(f;\bm{\lambda}) \:
        e^{-2\pi i f\, \Delta t(\alpha,\delta)} ,
\label{eq:detector-response-22}
\end{equation}
where
${
    {\cal F}(t;\alpha,\delta,\psi,\iota,D_L)
    \equiv
    F_+ C_+ + F_\times C_\times
}$
is the time-dependent complex detector-response amplitude multiplying the intrinsic
waveform.  Here
$
    C_ += \{1+\cos^2\iota\}/{2D_L},
$
and 
$
    {C_\times=-{i\,\cos\iota}/{D_L},}
$
${(\alpha, \delta)}$ are the geocentric sky coordinates, $\psi$ is the
polarization angle, $\iota$ is the inclination, and $D_L$ the
luminosity distance.  
For a source along $\hat{\bm n}$ and the vector $\bm r_d$ from geocenter to detector, 
${\Delta t(t;\alpha,\delta)
=\hat{\bm n}\!\cdot\!\bm r_d(t)/c}$ is the
geocenter-to-detector delay.  Its time dependence due to Earth's
rotation produces the rotational Doppler phase modulation.  The common
time argument of ${\cal F}$ and $\Delta t$ is left implicit here and
identified with the stationary time below.

\emph{\textbf{The long-signal inference barrier.}}---
For short signals, the detector orientation and position change
negligibly during the observation, so ${\cal F}$ and $\Delta t$ are
effectively constant.  The antenna response can then be pulled outside
the frequency-domain quantities that enter Bayesian parameter
estimation, namely the data--template correlation and the template
norm.  
These expensive frequency-domain calculations are then independent of
sky position, polarization, inclination, and distance; coalescence time
enters through a simple phase shift, so they can be precomputed or
compressed once while the Bayesian sampler explores the extrinsic parameters.



%
For a multi-hour inspiral, both the antenna patterns and the
geocenter-to-detector delay vary appreciably during the observation.
A direct rotating-response model therefore places sky-dependent
antenna factors inside both the frequency-domain data--template
correlation and the template norm.  The time-dependent delay
additionally contributes a sky-dependent rotational Doppler phase to
the correlation; being a pure phase, it cancels from the template
norm.  Without further reorganization, the usual separation between
expensive intrinsic calculations and cheap extrinsic projections is
lost as the sampler moves through the extrinsic parameter space.


The cost is compounded by the signal duration.  A GW170817-like binary
neutron star system~\cite{Abbott:2017xzu} remains in band for $\simeq2.8\,\mathrm{min}$ above
$20\,\mathrm{Hz}$ but $\simeq21.3\,\mathrm{h}$ above $2\,\mathrm{Hz}$.
On a duration-resolving grid, $\Delta f=1/T$, this increases the number
of positive-frequency samples by more than two orders of magnitude.
Long duration therefore makes each direct overlap expensive, while
Earth's rotation makes it extrinsic-parameter dependent.
\emph{This coupling is the long-signal barrier for 3G
gravitational-wave inference.}

Fast inference for long-duration signals has been explored using several
compression strategies.  Smith~\textit{et al.}~\cite{Smith2021}
demonstrated 3G BNS inference from 5~Hz using reduced-order quadrature
with Earth rotation, corresponding to signals lasting only
$\simeq1.9$~h for the binary considered here, while 
Guttman~\textit{et al.} extended reduced-order methods further into the 
long-duration regime~\cite{Guttman2026}.
Heterodyned likelihoods have enabled rapid inference for long LISA
signals~\cite{Katz2022,Hoy2024}, and relative binning has been extended
to more general waveforms and 3G applications~\cite{Narola2024,
KumarGuptaSathyaprakash2025}; compression nevertheless becomes
challenging toward a few hertz~\cite{Baker2025}.

Baral~\textit{et al.} included Earth rotation and frequency-dependent
response in single-CE inference~\cite{Baral2023,Baral2025}, but their
relative-binning demonstration used the injected waveform as the
fiducial.  Tenorio and Gerosa~\cite{TenorioGerosa2025} used the Jaranowski--Kr\'olak--Schutz (JKS)
decomposition for matched filtering.  Here we instead use its finite
harmonic structure to factorize the Bayesian likelihood, removing
sky-dependent antenna modulation from the expensive frequency-domain
summaries without prior knowledge of the source sky position.

In this \textit{Letter}, we address both the rotation-induced
extrinsic-parameter dependence and the large frequency-domain costs
separately.  The Jaranowski--Kr\'olak--Schutz representation
replaces the rotating antenna response by a fixed five-function basis,
requiring no prior knowledge of the source sky position.  A
curvature-controlled frequency grid compresses the expensive
frequency-domain calculations into reusable likelihood summaries,
while relative binning reduces each subsequent likelihood evaluation
to a sparse frequency grid, where the remaining waveform and Doppler
variations are applied.


%

\emph{\textbf{The stationary-time map.}}---
The antenna response and detector delay vary on the Earth-rotation
timescale and are slow compared with the binary's rapidly varying
orbital phase.  The usual stationary-phase treatment~\cite{Sathyaprakash:1991mt,Cutler:1994ys} can therefore be
applied with these slowly varying quantities evaluated at the
stationary time $\tf(f)$,
defined by $2\pi f = \dot\phi(\tf(f))$, where $\phi(t)$ is the phase of the
dominant $(2,2)$ waveform. For the Newtonian $(2,2)$ phase, inverting the
chirp-rate relation $\dot f_{\rm gw}\propto f_{\rm gw}^{11/3}$ gives $\tf(f)$
in closed form,
\begin{equation}
  \tf(f) = t_c - \frac{5}{256}\,\Mc^{-5/3}(\pi f)^{-8/3},
  \label{eq:tf-second}
\end{equation}
making explicit that $\tf(f)$ depends on the source only through its chirp
mass $\Mc$. Eq.\eqref{eq:tf-second} provides a useful Newtonian estimate of the
time--frequency scaling. 
In the numerical calculation presented later, however, the
stationary-time map is obtained once from the full waveform phase at
a fiducial point $\bm\lambda_0$, denoted $t_f^{\bm\lambda_0}(f)$, and then held fixed
during sampling.

\emph{\textbf{The JKS split}.---}
%
The Jaranowski--Kr\'olak--Schutz (JKS)
decomposition~\cite{JKS1998} gives a finite sidereal expansion of the
antenna response of a fixed ground-based interferometer in the
long-wavelength approximation.  For a source fixed on the sky, the
time dependence arises from Earth's rotation,
$\Phi(t)\equiv{\rm GMST}(t)$, which rigidly rotates the rank-two
detector tensor about Earth's spin axis.  The response therefore
contains only the zeroth, first, and second sidereal harmonics:
\begin{equation}
    {\cal F}(t;\alpha,\delta,\psi,\iota,D_L)
    =
    \sum_{n=1}^{5}
        G_n(\alpha,\delta,\psi,\iota,D_L)\: e_n(t),
\label{eq:jks-combined-response}
\end{equation}
with
$
    e_n(t)
    \in
    \{\cos2\Phi(t),\sin2\Phi(t),\cos\Phi(t),\sin\Phi(t),1\}.
\label{eq:jks_basis}
$
The size of this basis is independent of the signal duration.

The coefficients $G_n$ are algebraic functions of the extrinsic
parameters; their explicit form is not needed below.
The important point is that all sidereal time dependence is carried by
the five known functions $e_n(t)$, while the dependence on sky position,
polarization, inclination, and distance is carried by the five
coefficients $G_n$ for each detector.

\emph{\textbf{Intrinsic/extrinsic factorization.}}---
The JKS form~Eq.\eqref{eq:jks-combined-response} decouples the time dependence of the rotating response from its dependence on extrinsic parameters, replacing it with
five known sidereal functions multiplied by five extrinsic
coefficients. To make these functions reusable over a
local intrinsic domain, we evaluate them along a common
fiducial stationary-time map 
$
  e_n^{\bm\lambda_0}(f) =
        e_n[t_f^{\bm\lambda_0}(f)]
$. 
We therefore define
$
R_n(f;\bm\lambda)
\equiv
e_n^{\bm\lambda_0}(f)\, \tilde h_{22}(f;\bm\lambda).
$
so that the detector waveform becomes
\begin{equation}
  \tilde h(f)
  =
  \sum_{n=1}^{5}G_n \:
  R_n(f;\bm\lambda)\:
  \exp\left (-2\pi i f\:\Delta t[\alpha,\delta; t_f^{\bm\lambda_0}(f)]\right ) .
  \label{eq:factorized-response}
\end{equation}
%
The additional approximation introduced in Eq.\eqref{eq:factorized-response} is that the stationary-time
map entering the sidereal functions and detector delay is constructed
once from the full waveform phase at a fiducial intrinsic point
$\bm\lambda_0$ and is not re-evaluated as the
sampler proposes new intrinsic parameters.  The intrinsic waveform
$\tilde h_{22}(f;\bm{\lambda})$, including its phase, is
nevertheless evaluated at every trial point.

The size of this fixed-map approximation can be estimated using the
Newtonian expression Eq.\eqref{eq:tf-second}. 
This approximation is small because the stationary-time map changes
little over the local likelihood support.
A displacement in chirp mass changes the stationary time by
$
\Delta t_f(f) \propto (\mathcal{M}_{c,0})^{-5/3}
(\pi f)^{-8/3}\,\Delta\ln\mathcal{M}_c,
$
so the corresponding sidereal-phase change is
$
\Delta\Phi_{\rm sid}(f)
\simeq
\Omega_\oplus\,\Delta t_f(f).
$
Over a local MM$=0.95$ minimal-match ellipsoid, taken as a
representative likelihood-support domain, this gives
$\max|\Delta\Phi_{\rm sid}|=4.3\times10^{-5}$~rad, far below the
phase tolerances used below.

\emph{\textbf{Curvature-controlled grid and acceleration by relative
binning.}}---
We write the detector data as ${d=n+s}$, where $s$ is the signal present
in the data and $h$ denotes a trial waveform.  The two quantities entering
the phase-marginalized likelihood are the complex data--template
correlation $z(\tau)$ and the template norm,
\begin{equation}
\ln \mathcal L(\tau)
=
\ln I_0\!\left(|z(\tau)|\right)
-\frac{1}{2} \norm{h}^2.
\label{eq:phase_marg}
\end{equation}
Writing the rotational Doppler factor
${{\cal D}(f;\alpha,\delta)\equiv
\exp\{-2\pi i f\: \Delta t[\alpha,\delta;t_f^{\bm\lambda_0}(f)]}\}$
and defining ${z_n(\tau)\equiv\left\langle{\cal D}R_n|d \right\rangle_{\mathbb C}(\tau)}$,
the JKS representation gives
\begin{equation}
z(\tau)=\sum_{n=1}^{5}G_n^*z_n(\tau),
\quad
\norm{h}^2=\sum_{n,m=1}^{5}G_n^*G_mH_{nm},
\label{eq:z_and_norm}
\end{equation}
where $H_{nm}\equiv\langle R_n|R_m\rangle$ is the Gram matrix.
Here $\tau\equiv\Delta t_c^{\rm geo}$ is the geocentric
coalescence-time offset. Together with
the rotational delay in ${\cal D}$, the correlation therefore carries
the total phase
$\exp\{-2\pi i f\: (\tau \, +\, \Delta t[\alpha,\delta; t_f^{\bm\lambda_0}(f)])\}$.
Here $\langle\cdot|\cdot\rangle_{\mathbb C}$ denotes the complex
noise-weighted correlation, with $\langle\cdot|\cdot\rangle$ its real
part.  The phase factor enters $z_n$ but cancels from $H_{nm}$; its
efficient treatment is introduced below.  \textcolor{black}{The JKS split has
therefore removed the sky-dependent antenna amplitude from the
frequency sums; the remaining sky dependence in ${\cal D}$ is moved
to the sparse online evaluation below.}

We use a single fiducial intrinsic point $\bm\lambda_0$ throughout the
local construction. It defines the fixed fiducial stationary-time map entering the JKS
basis, the corresponding JKS-projected intrinsic waveforms $R_n$,
the reference waveform
$\tilde h_0(f)=\tilde h_{22}(f;\bm\lambda_0)$, and its fiducial phase
$\Psi_0(f)\equiv\Psi(f;\bm\lambda_0)$; the latter is used to construct
both frequency grids below.

The JKS split removes the antenna-amplitude dependence from the
frequency sums, but not their resolution cost.  We therefore construct
the smooth intrinsic objects on a uniform curvature-controlled grid,
\begin{equation}
|\Psi_0''(f_{\rm low})|\,\Delta f_{\rm curv}^{\,2}
\lesssim
\epsilon_{\rm grid},
\label{eq:curv_grid}
\end{equation}
with nominal $\epsilon_{\rm grid}=0.1$ rad.  This bounds the leading
quadratic phase remainder, while the coalescence-time and
detector-delay phases are inserted later at the relative-bin edges.
For the zero-noise example below, this reduces the reusable
construction from $N_{\rm native}$ to $N_{\rm curvature}$ samples.
%
%

\emph{\textbf{The online relative-binning layer.}---}
For each trial intrinsic
point $\bm\lambda$, define
\begin{equation}
\rho(f;\bm\lambda)
=
\frac{\tilde h_{22}(f;\bm\lambda)}
{\tilde h_{22}(f;\bm\lambda_0)},
\quad
r(f)=\rho(f;\bm\lambda)e^{-2\pi i f\, \Delta\tau(f)},
\label{eq:relbin_ratio_def}
\end{equation}
where
${\Delta\tau(f) =
\tau+\Delta t[\alpha,\delta;t_f^{\bm\lambda_0}(f)]}$
includes the coalescence-time offset and the time-dependent
geocenter-to-detector delay. The JKS split factorizes the rotating
antenna amplitude, while the remaining phase is incorporated into
$r(f)$ only at the relative-bin edges.

Within each bin $b$, centered at $f_b$, we approximate
$\rho(f)\simeq\rho_0^b+\rho_1^b(f-f_b)$ and
$r(f)\simeq r_0^b+r_1^b(f-f_b)$. The nonuniform bin widths are fixed by
${
\left[
|\Psi_0''(f_b)|
+
(2\pi\Delta\tau_{\max})^2
\right]\Delta f_b^2
\lesssim
\epsilon_b,
}$
%
%
where $\Delta\tau_{\max}$ bounds the allowed time shift.  As in
Eq.(\ref{eq:curv_grid}), this conservatively bounds the leading
quadratic remainder, including both intrinsic phase curvature and the
allowed coalescence-time and Doppler shifts.

As $\rho(f)$ and $r(f)$ are linearized within each bin, the
reference data--waveform correlations and waveform--waveform overlaps
entering Eq.\eqref{eq:z_and_norm} need only be compressed into their
zeroth- and first-order moments
$\mathcal A^{b}_{0,n}$, $\mathcal A^{b}_{1,n}$ (the $\mathcal A$
summaries) and
$\mathcal B^{b}_{0,nm}$, $\mathcal B^{b}_{1,nm}$ (the $\mathcal B$
summaries).
The phase-marginalized likelihood 
Eq.~\eqref{eq:phase_marg}, can then be evaluated online using
\begin{align}
z(\tau)
&\simeq
\sum_{b,n}G_n^*
\bigl(r_0^b\mathcal A_{0,n}^{b}
+r_1^b\mathcal A_{1,n}^{b}\bigr),
\label{eq:relbin_data_term}
\\
\norm{h}^2
&\simeq
\sum_{b,n,m}G_n^*G_m
\bigl(
|\rho_0^b|^2\mathcal B_{0,nm}^{b}
+\eta_b\, \mathcal B_{1,nm}^{b}
\bigr),
\label{eq:relbin_norm_term}
\end{align}
where $\eta_b\equiv 2\,{\rm Re}(\rho_0^{b*}\rho_1^b)$.
The phase shift enters only through $r_{0,1}^b$ and cancels from the
norm. Changing the extrinsic parameters therefore updates only the
coefficients $G_n$ and the phase factors evaluated at the bin edges,
without rebuilding the summaries.
\begin{figure}[t]
\centering
\includegraphics[width=\linewidth]{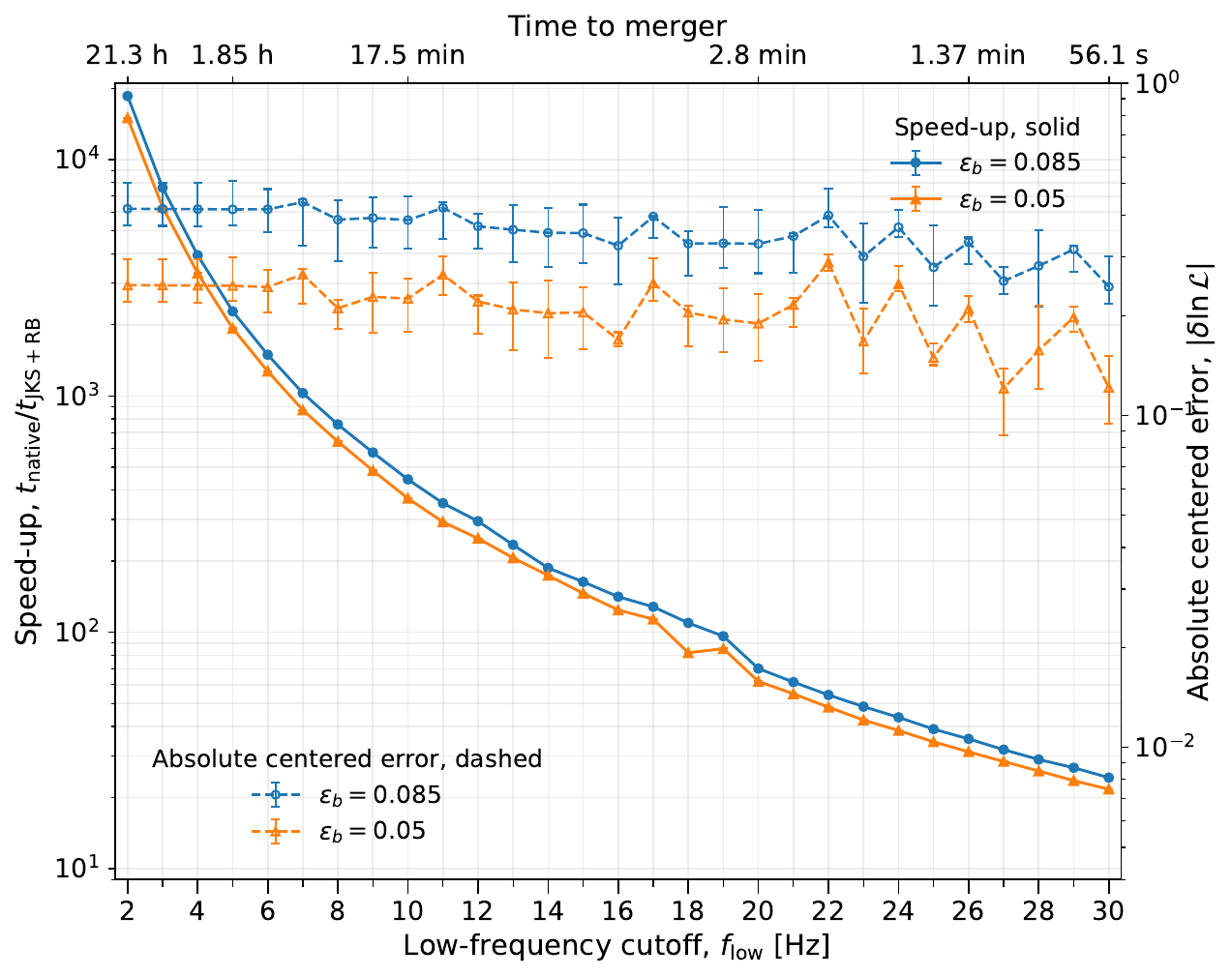}
\caption{Speed-up of the JKS-factored relative-binning likelihood over
the streamed native-resolution likelihood for a GW170817-like BNS with
H1 geometry and the ET-D PSD. Solid curves show median per-call timings
from repeated evaluations at the injection parameters; dashed curves
show the median absolute centered error $|\delta\ln\mathcal{L}|$ over 31
nearby validation points. Error bars give central 50\% bootstrap
intervals, and the upper axis gives the Newtonian time to merger. At
$f_{\rm low}=2\,\mathrm{Hz}$, the median errors are $0.416$ and $0.245$
for $\epsilon_b=0.085$ and $0.05$, respectively. We use
$\epsilon_b=0.05$ for the parameter-estimation runs below.}
\label{fig:speed_and_accuracy}
\end{figure}
For the BNS example specified below over $[2,1800]\,{\rm Hz}$, the
duration-resolving grid contains
$N_{\rm native}=1.383\times10^8$ positive-frequency samples,
whereas the curvature-controlled construction uses
$N_{\rm curvature}=8.379\times10^6$, a reduction by a factor
of $16.5$.  For the adopted choice $\epsilon_b=0.05$, the
online likelihood requires only
$M_{\rm bin}=1.0295\times10^4$ relative bins, giving
${
N_{\rm native}\gg N_{\rm curvature}\gg M_{\rm bin}.
}$
These ratios are not wall-clock speedups, since waveform evaluation and
JKS contractions also contribute; they distinguish the duration-grid
baseline, curvature-grid setup, and relative-bin online cost.

For zero-noise data, both $\mathcal A$ and $\mathcal B$ summaries may be
constructed on the curvature grid. For real data, nonsmooth noise
requires $\mathcal A$ to be accumulated once over the native Fourier
bins, while waveform-only $\mathcal B$ remains on the curvature grid.
The costs are therefore $\mathcal O(N_{\rm native})$ once for
$\mathcal A$, $\mathcal O(N_{\rm curvature})$ once for $\mathcal B$,
and $\mathcal O(M_{\rm bin})$ per likelihood call. Figure~1 uses
native-grid $\mathcal A$ and curvature-grid $\mathcal B$.


%

\begin{figure*}[t]
\centering
\includegraphics[width=0.625\linewidth]{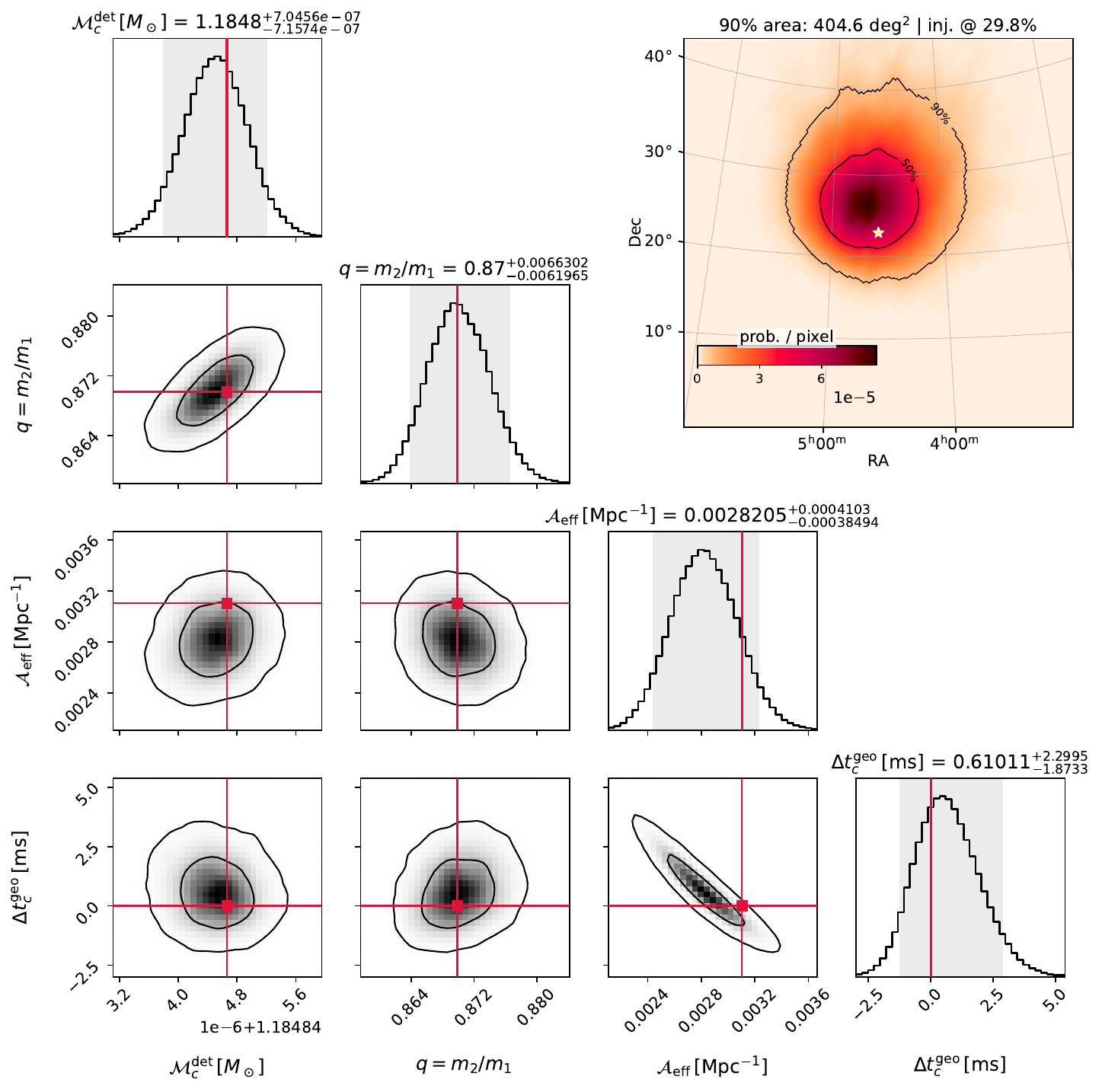}
\caption{Equal-run-weight pooled posterior from $50$ independent
\textsc{dynesty} runs for $\Mc^{\rm det}$, $q$, $A_{\rm eff}$, and
the geocentric coalescence-time offset $\Delta t_c^{\rm geo}$.
Diagonal titles give the median and $90\%$ credible interval, and red
lines mark the injected values.  The inset shows the corresponding
single-detector sky localization, with the $50\%$ and $90\%$ credible
regions and the injected sky position marked by a star.  The orientation angles $\iota$ and $\psi$ are not shown separately;
$A_{\rm eff}$ summarizes the distance--inclination amplitude
information measured most directly by the single-detector response.}
\label{fig:corner}
\end{figure*}

\emph{\textbf{Demonstration.}---}
%
We test the method on a non-spinning GW170817-like binary neutron star
with detector-frame component masses
$(m_1,m_2)=(1.46,1.27)\,M_\odot$, over the bandwidth
$[2,1800]\,{\rm Hz}$,
using zero-noise data, the ET-D sensitivity curve~\cite{Hild2011}, and H1 detector geometry~\cite{Aasi:2014mqd}.  
We sample the eight parameters
$(\Mc^{\rm det},q,\alpha,\delta,\psi,\iota,D_L,t_c)$.


\mbox{Fig.~(\ref{fig:speed_and_accuracy})} summarizes the speed and accuracy of the
JKS+relative-binning likelihood.  We define the centered
log-likelihood error as
${
\delta\ln\mathcal L(\bm\theta)
=
\Delta\ln\mathcal L(\bm\theta)
-
\Delta\ln\mathcal L(\bm\theta_{\rm inj})}$, where 
${
\Delta\ln\mathcal L
\equiv
\ln\mathcal L_{\rm JKS+RB}
-
\ln\mathcal L_{\rm native}.
}$
Comparing against the native-resolution likelihood at 31 nearby
parameter-space points, the adopted choice $\epsilon_b=0.05$ gives,
at $f_{\rm low}=2\,{\rm Hz}$, a median absolute centered
log-likelihood error of $|\delta\ln\mathcal L|=0.245$.

Relaxing the binning tolerance
to $\epsilon_b=0.085$ increases this to $0.416$, showing the expected
improvement as the bins are refined.  We emphasize that these are
absolute errors in $\ln\mathcal L$, not fractional or percentage
errors.  For $\epsilon_b=0.05$, the likelihood requires about
$2.6\,{\rm ms}$ per call on a single CPU core, compared with
$\simeq39\,{\rm s}$ for the streamed native-resolution likelihood at
$2\,{\rm Hz}$, corresponding to a speed-up of
$\simeq1.5\times10^4$.  The looser $\epsilon_b=0.085$ choice reaches
$\simeq1.86\times10^4$, at the cost of the larger likelihood error
shown in \mbox{Fig.~(\ref{fig:speed_and_accuracy})}.

Fig.~(\ref{fig:corner}) shows the eight-parameter inference obtained with the adopted
choice $\epsilon_b=0.050$, combining $50$ independent
\textsc{dynesty} runs with equal weight.  \textcolor{black}{The posterior is
stable across sampler seeds; an independent native-likelihood check
on posterior samples is given in the Supplemental Material.}  Apart from the weakly constrained
orientation angles $\iota$ and $\psi$, the largest displacement of an
individual run median from the ensemble median is $0.78$ times the
typical $16$--$84\%$ posterior half-width, and occurs for the
luminosity distance $D_L$.

Across $50$ independent \textsc{dynesty}~\cite{Speagle2020} runs
using random slice sampling (\texttt{rslice})~\cite{Neal2003}, 
the median sampling time is
$311.6\,{\rm s}$ on 64 cores for $3.58\times10^6$ likelihood
evaluations, with a weighted effective sample size of $1.39\times10^4$;
likelihood construction adds $11.2\,{\rm s}$.

For the single-detector amplitude information, we use
$
A_{\rm eff}\equiv
\sqrt{
\left(\frac{1+\cos^2\iota}{2}\right)^2+\cos^2\iota
}\ /D_L,
$
which captures the distance--inclination combination measured most
directly by the dominant-mode response. 
From the pooled posterior, the $68\%$ half-widths correspond to
precisions of $0.36$ ppm in $\Mc^{\rm det}$, $\simeq0.45\%$ in $q$,
$\simeq8.7\%$ in $A_{\rm eff}$, and $\simeq1.27\,{\rm ms}$
in the geocentric coalescence time.
%
Although not shown in
Fig.~(\ref{fig:corner}), the corresponding arrival time at H1 is measured much more
sharply, to $\simeq0.036\,{\rm ms}$.  The $90\%$ sky area is
$\mathrm{404.6}\,{\rm deg}^{2}$, with the injected position at the
$\mathrm{29.8}\%$ credible level.  The quoted uncertainties and sky
area characterize this representative single-detector H1 geometry
with the ET-D PSD, rather than a final ET or CE network forecast.

The calculation was performed on a single dual-socket workstation with
two 32-core AMD EPYC 7542 processors (64 physical cores total) and
512\,GB RAM.  Thus a BNS signal that remains in band for nearly a full
sidereal day can be analyzed over eight parameters in about five
minutes of sampling on one compute node, while retaining
single-detector sky and timing information.

\emph{\textbf{Conclusion.}---}
For the signals considered here, the long-signal PE barrier can be overcome 
by exposing the
finite harmonic structure of the rotating detector response.
The exact five-harmonic JKS decomposition removes the sky-dependent antenna
modulation from the expensive frequency-domain summaries, while the
rotational Doppler phase is retained in the sparse online likelihood.
A curvature-controlled grid separately reduces the cost of constructing
those summaries. The resulting likelihood requires no prior knowledge of
the source sky position and scales online with the relative-bin grid,
$\mathcal{O}(M_{\rm bin})$, rather than the native duration grid.

The five-harmonic factorization derived here is valid in the long-wavelength
limit and can be combined directly with other fast PE methods, including
meshfree/RBF likelihood surrogates~\cite{Pathak2023,Pathak2024,9xck-m23j}.
Extending this work to finite-arm effects in CE,
particularly its 40-km interferometer, where frequency and sky direction
couple and the exact five-term form is lost, is left for future work.

\begin{acknowledgments}
\emph{\textbf{Acknowledgments.}---}
I thank LSC colleagues Michael Williams, P. Ajith and Paul Lasky for useful comments and feedback. Thanks are due to LIGO-India Scientific Collaboration colleagues for helpful feedback. I also thank my graduate students Abhishek Sharma and Divya Tahelyani for helpful discussions.  I also thank IIT Gandhinagar for access to the {\texttt{tarang}} workstation on which most of the computations were carried out. Thanks are also due to IIT Gandhinagar's ISTF staff who provided prompt support with the computational infrastructure.
\end{acknowledgments}

\bibliography{refs}

\begin{thebibliography}{26}%
\makeatletter
\providecommand \@ifxundefined [1]{%
 \@ifx{#1\undefined}
}%
\providecommand \@ifnum [1]{%
 \ifnum #1\expandafter \@firstoftwo
 \else \expandafter \@secondoftwo
 \fi
}%
\providecommand \@ifx [1]{%
 \ifx #1\expandafter \@firstoftwo
 \else \expandafter \@secondoftwo
 \fi
}%
\providecommand \natexlab [1]{#1}%
\providecommand \enquote  [1]{``#1''}%
\providecommand \bibnamefont  [1]{#1}%
\providecommand \bibfnamefont [1]{#1}%
\providecommand \citenamefont [1]{#1}%
\providecommand \href@noop [0]{\@secondoftwo}%
\providecommand \href [0]{\begingroup \@sanitize@url \@href}%
\providecommand \@href[1]{\@@startlink{#1}\@@href}%
\providecommand \@@href[1]{\endgroup#1\@@endlink}%
\providecommand \@sanitize@url [0]{\catcode `\\12\catcode `\$12\catcode
  `\&12\catcode `\#12\catcode `\^12\catcode `\_12\catcode `\%12\relax}%
\providecommand \@@startlink[1]{}%
\providecommand \@@endlink[0]{}%
\providecommand \url  [0]{\begingroup\@sanitize@url \@url }%
\providecommand \@url [1]{\endgroup\@href {#1}{\urlprefix }}%
\providecommand \urlprefix  [0]{URL }%
\providecommand \Eprint [0]{\href }%
\providecommand \doibase [0]{https://doi.org/}%
\providecommand \selectlanguage [0]{\@gobble}%
\providecommand \bibinfo  [0]{\@secondoftwo}%
\providecommand \bibfield  [0]{\@secondoftwo}%
\providecommand \translation [1]{[#1]}%
\providecommand \BibitemOpen [0]{}%
\providecommand \bibitemStop [0]{}%
\providecommand \bibitemNoStop [0]{.\EOS\space}%
\providecommand \EOS [0]{\spacefactor3000\relax}%
\providecommand \BibitemShut  [1]{\csname bibitem#1\endcsname}%
\let\auto@bib@innerbib\@empty
\bibitem [{\citenamefont {Evans}\ \emph {et~al.}(2021)\citenamefont {Evans},
  \citenamefont {Adhikari}, \citenamefont {Afle}, \citenamefont {Ballmer} \emph
  {et~al.}}]{evans2021horizonstudycosmicexplorer}%
  \BibitemOpen
  \bibfield  {author} {\bibinfo {author} {\bibfnamefont {M.}~\bibnamefont
  {Evans}}, \bibinfo {author} {\bibfnamefont {R.~X.}\ \bibnamefont {Adhikari}},
  \bibinfo {author} {\bibfnamefont {C.}~\bibnamefont {Afle}}, \bibinfo {author}
  {\bibfnamefont {S.~W.}\ \bibnamefont {Ballmer}}, \emph {et~al.},\ }\href
  {https://arxiv.org/abs/2109.09882} {\bibinfo {title} {A horizon study for
  cosmic explorer: Science, observatories, and community}} (\bibinfo {year}
  {2021}),\ \Eprint {https://arxiv.org/abs/2109.09882} {arXiv:2109.09882
  [astro-ph.IM]} \BibitemShut {NoStop}%
\bibitem [{\citenamefont {Key}\ \emph {et~al.}(2026)\citenamefont {Key},
  \citenamefont {Han},\ and\ \citenamefont {Ara}}]{Key_2026}%
  \BibitemOpen
  \bibfield  {author} {\bibinfo {author} {\bibfnamefont {J.~S.}\ \bibnamefont
  {Key}}, \bibinfo {author} {\bibfnamefont {R.~J.}\ \bibnamefont {Han}},\ and\
  \bibinfo {author} {\bibfnamefont {S.~R.}\ \bibnamefont {Ara}} (\bibinfo
  {collaboration} {Cosmic Explorer Project}),\ }\bibfield  {title} {\bibinfo
  {title} {{Cosmic Explorer Observatory Conceptual Design}},\ }\href
  {https://doi.org/10.1088/1742-6596/3177/1/012098} {\bibfield  {journal}
  {\bibinfo  {journal} {Journal of Physics: Conference Series}\ }\textbf
  {\bibinfo {volume} {3177}},\ \bibinfo {pages} {012098} (\bibinfo {year}
  {2026})}\BibitemShut {NoStop}%
\bibitem [{\citenamefont {Maggiore}\ \emph {et~al.}(2020)\citenamefont
  {Maggiore} \emph {et~al.}}]{Maggiore:2019uih}%
  \BibitemOpen
  \bibfield  {author} {\bibinfo {author} {\bibfnamefont {M.}~\bibnamefont
  {Maggiore}} \emph {et~al.} (\bibinfo {collaboration} {ET}),\ }\bibfield
  {title} {\bibinfo {title} {{Science Case for the Einstein Telescope}},\
  }\href {https://doi.org/10.1088/1475-7516/2020/03/050} {\bibfield  {journal}
  {\bibinfo  {journal} {JCAP}\ }\textbf {\bibinfo {volume} {03}},\ \bibinfo
  {pages} {050}},\ \Eprint {https://arxiv.org/abs/1912.02622} {arXiv:1912.02622
  [astro-ph.CO]} \BibitemShut {NoStop}%
\bibitem [{\citenamefont {Aasi}\ \emph {et~al.}(2015)\citenamefont {Aasi} \emph
  {et~al.}}]{Aasi:2014mqd}%
  \BibitemOpen
  \bibfield  {author} {\bibinfo {author} {\bibfnamefont {J.}~\bibnamefont
  {Aasi}} \emph {et~al.} (\bibinfo {collaboration} {LIGO Scientific
  Collaboration}),\ }\bibfield  {title} {\bibinfo {title} {{Advanced LIGO}},\
  }\href {https://doi.org/10.1088/0264-9381/32/7/074001} {\bibfield  {journal}
  {\bibinfo  {journal} {Class. Quant. Grav.}\ }\textbf {\bibinfo {volume}
  {32}},\ \bibinfo {pages} {074001} (\bibinfo {year} {2015})},\ \Eprint
  {https://arxiv.org/abs/1411.4547} {arXiv:1411.4547 [gr-qc]} \BibitemShut
  {NoStop}%
\bibitem [{\citenamefont {Acernese}\ \emph {et~al.}(2015)\citenamefont
  {Acernese} \emph {et~al.}}]{Acernese:2014hva}%
  \BibitemOpen
  \bibfield  {author} {\bibinfo {author} {\bibfnamefont {F.}~\bibnamefont
  {Acernese}} \emph {et~al.} (\bibinfo {collaboration} {Virgo Collaboration}),\
  }\bibfield  {title} {\bibinfo {title} {{Advanced Virgo: a second-generation
  interferometric gravitational wave detector}},\ }\href
  {https://doi.org/10.1088/0264-9381/32/2/024001} {\bibfield  {journal}
  {\bibinfo  {journal} {Class. Quant. Grav.}\ }\textbf {\bibinfo {volume}
  {32}},\ \bibinfo {pages} {024001} (\bibinfo {year} {2015})},\ \Eprint
  {https://arxiv.org/abs/1408.3978} {arXiv:1408.3978 [gr-qc]} \BibitemShut
  {NoStop}%
\bibitem [{\citenamefont {Akutsu}\ \emph {et~al.}(2021)\citenamefont {Akutsu}
  \emph {et~al.}}]{Akutsu:2020his}%
  \BibitemOpen
  \bibfield  {author} {\bibinfo {author} {\bibfnamefont {T.}~\bibnamefont
  {Akutsu}} \emph {et~al.} (\bibinfo {collaboration} {KAGRA Collaboration}),\
  }\bibfield  {title} {\bibinfo {title} {{Overview of KAGRA: Detector design
  and construction history}},\ }\href {https://doi.org/10.1093/ptep/ptaa125}
  {\bibfield  {journal} {\bibinfo  {journal} {Prog. Theor. Exp. Phys.}\
  }\textbf {\bibinfo {volume} {2021}},\ \bibinfo {pages} {05A101} (\bibinfo
  {year} {2021})}\BibitemShut {NoStop}%
\bibitem [{\citenamefont {Abbott}\ \emph {et~al.}(2017)\citenamefont {Abbott}
  \emph {et~al.}}]{Abbott:2017xzu}%
  \BibitemOpen
  \bibfield  {author} {\bibinfo {author} {\bibfnamefont {B.~P.}\ \bibnamefont
  {Abbott}} \emph {et~al.} (\bibinfo {collaboration} {LIGO Scientific
  Collaboration and Virgo Collaboration}),\ }\bibfield  {title} {\bibinfo
  {title} {{{GW170817: Observation of Gravitational Waves from a Binary Neutron
  Star Inspiral}}},\ }\href {https://doi.org/10.1103/PhysRevLett.119.161101}
  {\bibfield  {journal} {\bibinfo  {journal} {Phys. Rev. Lett.}\ }\textbf
  {\bibinfo {volume} {119}},\ \bibinfo {pages} {161101} (\bibinfo {year}
  {2017})},\ \Eprint {https://arxiv.org/abs/1710.05832} {arXiv:1710.05832
  [gr-qc]} \BibitemShut {NoStop}%
\bibitem [{\citenamefont {Smith}\ \emph {et~al.}(2021)\citenamefont {Smith},
  \citenamefont {Borhanian}, \citenamefont {Sathyaprakash}, \citenamefont
  {Hernandez~Vivanco}, \citenamefont {Field}, \citenamefont {Lasky},
  \citenamefont {Mandel}, \citenamefont {Morisaki}, \citenamefont {Ottaway},
  \citenamefont {Slagmolen}, \citenamefont {Thrane}, \citenamefont
  {T{\"o}yr{\"a}},\ and\ \citenamefont {Vitale}}]{Smith2021}%
  \BibitemOpen
  \bibfield  {author} {\bibinfo {author} {\bibfnamefont {R.}~\bibnamefont
  {Smith}}, \bibinfo {author} {\bibfnamefont {S.}~\bibnamefont {Borhanian}},
  \bibinfo {author} {\bibfnamefont {B.~S.}\ \bibnamefont {Sathyaprakash}},
  \bibinfo {author} {\bibfnamefont {F.}~\bibnamefont {Hernandez~Vivanco}},
  \bibinfo {author} {\bibfnamefont {S.~E.}\ \bibnamefont {Field}}, \bibinfo
  {author} {\bibfnamefont {P.~D.}\ \bibnamefont {Lasky}}, \bibinfo {author}
  {\bibfnamefont {I.}~\bibnamefont {Mandel}}, \bibinfo {author} {\bibfnamefont
  {S.}~\bibnamefont {Morisaki}}, \bibinfo {author} {\bibfnamefont
  {D.}~\bibnamefont {Ottaway}}, \bibinfo {author} {\bibfnamefont {B.~J.~J.}\
  \bibnamefont {Slagmolen}}, \bibinfo {author} {\bibfnamefont {E.}~\bibnamefont
  {Thrane}}, \bibinfo {author} {\bibfnamefont {D.}~\bibnamefont
  {T{\"o}yr{\"a}}},\ and\ \bibinfo {author} {\bibfnamefont {S.}~\bibnamefont
  {Vitale}},\ }\bibfield  {title} {\bibinfo {title} {Bayesian inference for
  gravitational waves from binary neutron star mergers in third-generation
  observatories},\ }\href {https://doi.org/10.1103/PhysRevLett.127.081102}
  {\bibfield  {journal} {\bibinfo  {journal} {Phys. Rev. Lett.}\ }\textbf
  {\bibinfo {volume} {127}},\ \bibinfo {pages} {081102} (\bibinfo {year}
  {2021})},\ \Eprint {https://arxiv.org/abs/2103.12274} {arXiv:2103.12274
  [gr-qc]} \BibitemShut {NoStop}%
\bibitem [{\citenamefont {Guttman}\ \emph {et~al.}(2026)\citenamefont
  {Guttman}, \citenamefont {Baker}, \citenamefont {Lasky},\ and\ \citenamefont
  {Thrane}}]{Guttman2026}%
  \BibitemOpen
  \bibfield  {author} {\bibinfo {author} {\bibfnamefont {N.}~\bibnamefont
  {Guttman}}, \bibinfo {author} {\bibfnamefont {A.~M.}\ \bibnamefont {Baker}},
  \bibinfo {author} {\bibfnamefont {P.~D.}\ \bibnamefont {Lasky}},\ and\
  \bibinfo {author} {\bibfnamefont {E.}~\bibnamefont {Thrane}},\ }\bibfield
  {title} {\bibinfo {title} {Licence to bin: Accurate and scalable inference
  for binary neutron stars in next-generation gravitational-wave detectors},\
  }\href@noop {} {\bibfield  {journal} {\bibinfo  {journal} {arXiv e-prints}\ }
  (\bibinfo {year} {2026})},\ \Eprint {https://arxiv.org/abs/2606.14197}
  {2606.14197} \BibitemShut {NoStop}%
\bibitem [{\citenamefont {Katz}(2022)}]{Katz2022}%
  \BibitemOpen
  \bibfield  {author} {\bibinfo {author} {\bibfnamefont {M.~L.}\ \bibnamefont
  {Katz}},\ }\bibfield  {title} {\bibinfo {title} {Fully automated end-to-end
  pipeline for massive black hole binary signal extraction from {LISA} data},\
  }\href {https://doi.org/10.1103/PhysRevD.105.044055} {\bibfield  {journal}
  {\bibinfo  {journal} {Phys. Rev. D}\ }\textbf {\bibinfo {volume} {105}},\
  \bibinfo {pages} {044055} (\bibinfo {year} {2022})},\ \Eprint
  {https://arxiv.org/abs/2111.01064} {arXiv:2111.01064 [gr-qc]} \BibitemShut
  {NoStop}%
\bibitem [{\citenamefont {Hoy}\ \emph {et~al.}(2024)\citenamefont {Hoy},
  \citenamefont {Weaving}, \citenamefont {Nuttall},\ and\ \citenamefont
  {Harry}}]{Hoy2024}%
  \BibitemOpen
  \bibfield  {author} {\bibinfo {author} {\bibfnamefont {C.}~\bibnamefont
  {Hoy}}, \bibinfo {author} {\bibfnamefont {C.}~\bibnamefont {Weaving}},
  \bibinfo {author} {\bibfnamefont {L.~K.}\ \bibnamefont {Nuttall}},\ and\
  \bibinfo {author} {\bibfnamefont {I.}~\bibnamefont {Harry}},\ }\bibfield
  {title} {\bibinfo {title} {A rapid multi-modal parameter estimation technique
  for {LISA}},\ }\href {https://doi.org/10.1088/1361-6382/ad8f26} {\bibfield
  {journal} {\bibinfo  {journal} {Class. Quantum Grav.}\ }\textbf {\bibinfo
  {volume} {41}},\ \bibinfo {pages} {245012} (\bibinfo {year} {2024})},\
  \Eprint {https://arxiv.org/abs/2408.12764} {arXiv:2408.12764 [gr-qc]}
  \BibitemShut {NoStop}%
\bibitem [{\citenamefont {Narola}\ \emph {et~al.}(2024)\citenamefont {Narola},
  \citenamefont {Janquart}, \citenamefont {Meijer}, \citenamefont {Haris},\
  and\ \citenamefont {Van Den~Broeck}}]{Narola2024}%
  \BibitemOpen
  \bibfield  {author} {\bibinfo {author} {\bibfnamefont {H.}~\bibnamefont
  {Narola}}, \bibinfo {author} {\bibfnamefont {J.}~\bibnamefont {Janquart}},
  \bibinfo {author} {\bibfnamefont {Q.}~\bibnamefont {Meijer}}, \bibinfo
  {author} {\bibfnamefont {K.}~\bibnamefont {Haris}},\ and\ \bibinfo {author}
  {\bibfnamefont {C.}~\bibnamefont {Van Den~Broeck}},\ }\bibfield  {title}
  {\bibinfo {title} {Gravitational-wave parameter estimation with relative
  binning},\ }\href {https://doi.org/10.1103/PhysRevD.110.084085} {\bibfield
  {journal} {\bibinfo  {journal} {Phys. Rev. D}\ }\textbf {\bibinfo {volume}
  {110}},\ \bibinfo {pages} {084085} (\bibinfo {year} {2024})},\ \Eprint
  {https://arxiv.org/abs/2308.12140} {arXiv:2308.12140 [gr-qc]} \BibitemShut
  {NoStop}%
\bibitem [{\citenamefont {Kumar}\ \emph {et~al.}(2025)\citenamefont {Kumar},
  \citenamefont {Gupta},\ and\ \citenamefont
  {Sathyaprakash}}]{KumarGuptaSathyaprakash2025}%
  \BibitemOpen
  \bibfield  {author} {\bibinfo {author} {\bibfnamefont {D.}~\bibnamefont
  {Kumar}}, \bibinfo {author} {\bibfnamefont {I.}~\bibnamefont {Gupta}},\ and\
  \bibinfo {author} {\bibfnamefont {B.}~\bibnamefont {Sathyaprakash}},\
  }\bibfield  {title} {\bibinfo {title} {Accelerating parameter estimation for
  parameterized tests of general relativity with gravitational-wave
  observations},\ }\href@noop {} {\  (\bibinfo {year} {2025})},\ \Eprint
  {https://arxiv.org/abs/2511.16879} {arXiv:2511.16879 [gr-qc]} \BibitemShut
  {NoStop}%
\bibitem [{\citenamefont {Baker}\ \emph {et~al.}(2025)\citenamefont {Baker},
  \citenamefont {Lasky}, \citenamefont {Thrane},\ and\ \citenamefont
  {Golomb}}]{Baker2025}%
  \BibitemOpen
  \bibfield  {author} {\bibinfo {author} {\bibfnamefont {A.~M.}\ \bibnamefont
  {Baker}}, \bibinfo {author} {\bibfnamefont {P.~D.}\ \bibnamefont {Lasky}},
  \bibinfo {author} {\bibfnamefont {E.}~\bibnamefont {Thrane}},\ and\ \bibinfo
  {author} {\bibfnamefont {J.}~\bibnamefont {Golomb}},\ }\bibfield  {title}
  {\bibinfo {title} {Significant challenges for astrophysical inference with
  next-generation gravitational-wave observatories},\ }\href@noop {} {\bibfield
   {journal} {\bibinfo  {journal} {Phys. Rev. D}\ }\textbf {\bibinfo {volume}
  {112}},\ \bibinfo {pages} {102004} (\bibinfo {year} {2025})}\BibitemShut
  {NoStop}%
\bibitem [{\citenamefont {Baral}\ \emph {et~al.}(2023)\citenamefont {Baral},
  \citenamefont {Morisaki}, \citenamefont {Maga{\~n}a~Hernandez},\ and\
  \citenamefont {Creighton}}]{Baral2023}%
  \BibitemOpen
  \bibfield  {author} {\bibinfo {author} {\bibfnamefont {P.}~\bibnamefont
  {Baral}}, \bibinfo {author} {\bibfnamefont {S.}~\bibnamefont {Morisaki}},
  \bibinfo {author} {\bibfnamefont {I.}~\bibnamefont {Maga{\~n}a~Hernandez}},\
  and\ \bibinfo {author} {\bibfnamefont {J.~D.~E.}\ \bibnamefont {Creighton}},\
  }\bibfield  {title} {\bibinfo {title} {Localization of binary neutron star
  mergers with a single {Cosmic Explorer}},\ }\href
  {https://doi.org/10.1103/PhysRevD.108.043010} {\bibfield  {journal} {\bibinfo
   {journal} {Phys. Rev. D}\ }\textbf {\bibinfo {volume} {108}},\ \bibinfo
  {pages} {043010} (\bibinfo {year} {2023})},\ \Eprint
  {https://arxiv.org/abs/2304.09889} {arXiv:2304.09889 [astro-ph.HE]}
  \BibitemShut {NoStop}%
\bibitem [{\citenamefont {Baral}\ \emph {et~al.}(2025)\citenamefont {Baral},
  \citenamefont {Morisaki}, \citenamefont {Gupta},\ and\ \citenamefont
  {Creighton}}]{Baral2025}%
  \BibitemOpen
  \bibfield  {author} {\bibinfo {author} {\bibfnamefont {P.}~\bibnamefont
  {Baral}}, \bibinfo {author} {\bibfnamefont {S.}~\bibnamefont {Morisaki}},
  \bibinfo {author} {\bibfnamefont {I.}~\bibnamefont {Gupta}},\ and\ \bibinfo
  {author} {\bibfnamefont {J.}~\bibnamefont {Creighton}},\ }\bibfield  {title}
  {\bibinfo {title} {Parameter estimation of gravitational-wave signals with
  frequency-dependent antenna responses and higher modes},\ }\href
  {https://doi.org/10.1088/1361-6382/ae128b} {\bibfield  {journal} {\bibinfo
  {journal} {Class. Quantum Grav.}\ }\textbf {\bibinfo {volume} {42}},\
  \bibinfo {pages} {215007} (\bibinfo {year} {2025})},\ \Eprint
  {https://arxiv.org/abs/2503.09627} {arXiv:2503.09627 [gr-qc]} \BibitemShut
  {NoStop}%
\bibitem [{\citenamefont {Tenorio}\ and\ \citenamefont
  {Gerosa}(2025)}]{TenorioGerosa2025}%
  \BibitemOpen
  \bibfield  {author} {\bibinfo {author} {\bibfnamefont {R.}~\bibnamefont
  {Tenorio}}\ and\ \bibinfo {author} {\bibfnamefont {D.}~\bibnamefont
  {Gerosa}},\ }\bibfield  {title} {\bibinfo {title} {Scalable data-analysis
  framework for long-duration gravitational waves from compact binaries using
  short fourier transforms},\ }\href
  {https://doi.org/10.1103/PhysRevD.111.104044} {\bibfield  {journal} {\bibinfo
   {journal} {Phys. Rev. D}\ }\textbf {\bibinfo {volume} {111}},\ \bibinfo
  {pages} {104044} (\bibinfo {year} {2025})},\ \Eprint
  {https://arxiv.org/abs/2502.11823} {arXiv:2502.11823 [gr-qc]} \BibitemShut
  {NoStop}%
\bibitem [{\citenamefont {Sathyaprakash}\ and\ \citenamefont
  {Dhurandhar}(1991)}]{Sathyaprakash:1991mt}%
  \BibitemOpen
  \bibfield  {author} {\bibinfo {author} {\bibfnamefont {B.~S.}\ \bibnamefont
  {Sathyaprakash}}\ and\ \bibinfo {author} {\bibfnamefont {S.~V.}\ \bibnamefont
  {Dhurandhar}},\ }\bibfield  {title} {\bibinfo {title} {{Choice of filters for
  the detection of gravitational waves from coalescing binaries}},\ }\href
  {https://doi.org/10.1103/PhysRevD.44.3819} {\bibfield  {journal} {\bibinfo
  {journal} {Phys. Rev. D}\ }\textbf {\bibinfo {volume} {44}},\ \bibinfo
  {pages} {3819} (\bibinfo {year} {1991})}\BibitemShut {NoStop}%
\bibitem [{\citenamefont {Cutler}\ and\ \citenamefont
  {Flanagan}(1994)}]{Cutler:1994ys}%
  \BibitemOpen
  \bibfield  {author} {\bibinfo {author} {\bibfnamefont {C.}~\bibnamefont
  {Cutler}}\ and\ \bibinfo {author} {\bibfnamefont {E.~E.}\ \bibnamefont
  {Flanagan}},\ }\bibfield  {title} {\bibinfo {title} {{Gravitational waves
  from merging compact binaries: How accurately can one extract the binary's
  parameters from the inspiral waveform?}},\ }\href
  {https://doi.org/10.1103/PhysRevD.49.2658} {\bibfield  {journal} {\bibinfo
  {journal} {Phys. Rev. D}\ }\textbf {\bibinfo {volume} {49}},\ \bibinfo
  {pages} {2658} (\bibinfo {year} {1994})},\ \Eprint
  {https://arxiv.org/abs/gr-qc/9402014} {arXiv:gr-qc/9402014} \BibitemShut
  {NoStop}%
\bibitem [{\citenamefont {Jaranowski}\ \emph {et~al.}(1998)\citenamefont
  {Jaranowski}, \citenamefont {Kr\'olak},\ and\ \citenamefont
  {Schutz}}]{JKS1998}%
  \BibitemOpen
  \bibfield  {author} {\bibinfo {author} {\bibfnamefont {P.}~\bibnamefont
  {Jaranowski}}, \bibinfo {author} {\bibfnamefont {A.}~\bibnamefont
  {Kr\'olak}},\ and\ \bibinfo {author} {\bibfnamefont {B.~F.}\ \bibnamefont
  {Schutz}},\ }\bibfield  {title} {\bibinfo {title} {{Data analysis of
  gravitational-wave signals from spinning neutron stars: The signal and its
  detection}},\ }\href {https://doi.org/10.1103/PhysRevD.58.063001} {\bibfield
  {journal} {\bibinfo  {journal} {Phys. Rev. D}\ }\textbf {\bibinfo {volume}
  {58}},\ \bibinfo {pages} {063001} (\bibinfo {year} {1998})}\BibitemShut
  {NoStop}%
\bibitem [{\citenamefont {Hild}\ \emph {et~al.}(2011)\citenamefont {Hild},
  \citenamefont {Abernathy}, \citenamefont {Acernese}, \citenamefont
  {Amaro-Seoane}, \citenamefont {Andersson}, \citenamefont {Arun} \emph
  {et~al.}}]{Hild2011}%
  \BibitemOpen
  \bibfield  {author} {\bibinfo {author} {\bibfnamefont {S.}~\bibnamefont
  {Hild}}, \bibinfo {author} {\bibfnamefont {M.}~\bibnamefont {Abernathy}},
  \bibinfo {author} {\bibfnamefont {F.}~\bibnamefont {Acernese}}, \bibinfo
  {author} {\bibfnamefont {P.}~\bibnamefont {Amaro-Seoane}}, \bibinfo {author}
  {\bibfnamefont {N.}~\bibnamefont {Andersson}}, \bibinfo {author}
  {\bibfnamefont {K.~G.}\ \bibnamefont {Arun}}, \emph {et~al.},\ }\bibfield
  {title} {\bibinfo {title} {Sensitivity studies for third-generation
  gravitational wave observatories},\ }\href
  {https://doi.org/10.1088/0264-9381/28/9/094013} {\bibfield  {journal}
  {\bibinfo  {journal} {Class. Quantum Grav.}\ }\textbf {\bibinfo {volume}
  {28}},\ \bibinfo {pages} {094013} (\bibinfo {year} {2011})},\ \Eprint
  {https://arxiv.org/abs/1012.0908} {arXiv:1012.0908 [gr-qc]} \BibitemShut
  {NoStop}%
\bibitem [{\citenamefont {Speagle}(2020)}]{Speagle2020}%
  \BibitemOpen
  \bibfield  {author} {\bibinfo {author} {\bibfnamefont {J.~S.}\ \bibnamefont
  {Speagle}},\ }\bibfield  {title} {\bibinfo {title} {\texttt{dynesty}: a
  dynamic nested sampling package for estimating bayesian posteriors and
  evidences},\ }\href {https://doi.org/10.1093/mnras/staa278} {\bibfield
  {journal} {\bibinfo  {journal} {Mon. Not. R. Astron. Soc.}\ }\textbf
  {\bibinfo {volume} {493}},\ \bibinfo {pages} {3132} (\bibinfo {year}
  {2020})}\BibitemShut {NoStop}%
\bibitem [{\citenamefont {Neal}(2003)}]{Neal2003}%
  \BibitemOpen
  \bibfield  {author} {\bibinfo {author} {\bibfnamefont {R.~M.}\ \bibnamefont
  {Neal}},\ }\bibfield  {title} {\bibinfo {title} {Slice sampling},\ }\href
  {https://doi.org/10.1214/aos/1056562461} {\bibfield  {journal} {\bibinfo
  {journal} {The Annals of Statististics}\ }\textbf {\bibinfo {volume} {31}},\
  \bibinfo {pages} {705} (\bibinfo {year} {2003})}\BibitemShut {NoStop}%
\bibitem [{\citenamefont {Pathak}\ \emph {et~al.}(2023)\citenamefont {Pathak},
  \citenamefont {Reza},\ and\ \citenamefont {Sengupta}}]{Pathak2023}%
  \BibitemOpen
  \bibfield  {author} {\bibinfo {author} {\bibfnamefont {L.}~\bibnamefont
  {Pathak}}, \bibinfo {author} {\bibfnamefont {A.}~\bibnamefont {Reza}},\ and\
  \bibinfo {author} {\bibfnamefont {A.~S.}\ \bibnamefont {Sengupta}},\
  }\bibfield  {title} {\bibinfo {title} {Fast likelihood evaluation using
  meshfree approximations for reconstructing compact binary sources},\ }\href
  {https://doi.org/10.1103/PhysRevD.108.064055} {\bibfield  {journal} {\bibinfo
   {journal} {Phys. Rev. D}\ }\textbf {\bibinfo {volume} {108}},\ \bibinfo
  {pages} {064055} (\bibinfo {year} {2023})},\ \Eprint
  {https://arxiv.org/abs/2210.02706} {arXiv:2210.02706 [gr-qc]} \BibitemShut
  {NoStop}%
\bibitem [{\citenamefont {Pathak}\ \emph {et~al.}(2024)\citenamefont {Pathak},
  \citenamefont {Munishwar}, \citenamefont {Reza},\ and\ \citenamefont
  {Sengupta}}]{Pathak2024}%
  \BibitemOpen
  \bibfield  {author} {\bibinfo {author} {\bibfnamefont {L.}~\bibnamefont
  {Pathak}}, \bibinfo {author} {\bibfnamefont {S.}~\bibnamefont {Munishwar}},
  \bibinfo {author} {\bibfnamefont {A.}~\bibnamefont {Reza}},\ and\ \bibinfo
  {author} {\bibfnamefont {A.~S.}\ \bibnamefont {Sengupta}},\ }\bibfield
  {title} {\bibinfo {title} {Prompt sky localization of compact binary sources
  using a meshfree approximation},\ }\href
  {https://doi.org/10.1103/PhysRevD.109.024053} {\bibfield  {journal} {\bibinfo
   {journal} {Phys. Rev. D}\ }\textbf {\bibinfo {volume} {109}},\ \bibinfo
  {pages} {024053} (\bibinfo {year} {2024})},\ \Eprint
  {https://arxiv.org/abs/2309.07012} {arXiv:2309.07012 [gr-qc]} \BibitemShut
  {NoStop}%
\bibitem [{\citenamefont {Sharma}\ \emph {et~al.}(2026)\citenamefont {Sharma},
  \citenamefont {Pathak}, \citenamefont {Roy},\ and\ \citenamefont
  {Sengupta}}]{9xck-m23j}%
  \BibitemOpen
  \bibfield  {author} {\bibinfo {author} {\bibfnamefont {A.}~\bibnamefont
  {Sharma}}, \bibinfo {author} {\bibfnamefont {L.}~\bibnamefont {Pathak}},
  \bibinfo {author} {\bibfnamefont {S.}~\bibnamefont {Roy}},\ and\ \bibinfo
  {author} {\bibfnamefont {A.~S.}\ \bibnamefont {Sengupta}},\ }\bibfield
  {title} {\bibinfo {title} {Rapid parameter estimation with the full symphony
  of compact binary mergers using a meshfree approximation},\ }\href
  {https://doi.org/10.1103/9xck-m23j} {\bibfield  {journal} {\bibinfo
  {journal} {Phys. Rev. D}\ }\textbf {\bibinfo {volume} {114}},\ \bibinfo
  {pages} {044043} (\bibinfo {year} {2026})},\ \Eprint
  {https://arxiv.org/abs/2508.04172} {arXiv:2508.04172 [gr-qc]} \BibitemShut
  {NoStop}%
\end{thebibliography}%

\clearpage

\linespread{0.98}\selectfont

\setcounter{section}{0}
\setcounter{equation}{0}
\setcounter{figure}{0}
\setcounter{table}{0}
\renewcommand{\thesection}{S\arabic{section}}
\renewcommand{\theequation}{S\arabic{equation}}
\renewcommand{\thefigure}{S\arabic{figure}}
\renewcommand{\thetable}{S\arabic{table}}

\renewcommand{\theHsection}{S\arabic{section}}
\renewcommand{\theHequation}{S\arabic{equation}}
\renewcommand{\theHfigure}{S\arabic{figure}}
\renewcommand{\theHtable}{S\arabic{table}}

\begin{center}
{\large\bfseries Supplemental Material for\\[0.25em]
``Fast Bayesian Inference for Long-Duration Gravitational-Wave Signals in 3G detectors''}\\[0.75em]
Anand S. Sengupta\\
Indian Institute of Technology Gandhinagar, Gandhinagar 382055, India
\end{center}

\vspace{0.5em}

This Supplemental Material follows the construction and numerical
demonstration in the Letter.  Section~\ref{sec:jks} expands the
five-harmonic Jaranowski--Kr\'olak--Schutz (JKS) response and its geometric
content; Sec.~\ref{sec:likelihood-construction} gives implementation details
and local accuracy checks for the fixed stationary-time map and accelerated
likelihood; Sec.~\ref{sec:posterior-native} gives the posterior-support
native-resolution validation referred to in the Letter; and
Sec.~\ref{sec:pe-details} records the parameter-estimation configuration and
sampler-seed stability underlying the pooled posterior.

\section{Five-harmonic detector response}
\label{sec:jks}

This section expands the discussion in \emph{The JKS split} and
\emph{Intrinsic/extrinsic factorization} in the Letter.  We make explicit
the geometric content of the coefficients $A_n$ and $B_n$ and validate the
five-harmonic reconstruction against the direct detector response.  The
decomposition is that of Jaranowski, Kr\'olak, and
Schutz~\cite{JKS1998}.

The quantities $A_n$ and $B_n$ used in the Letter are fixed geometric
coefficients.  Choose an Earth-fixed Cartesian basis
$(\hat{\bm X},\hat{\bm Y},\hat{\bm Z})$, with
$\hat{\bm Z}$ along the Earth's rotation axis.  For a detector at geodetic
latitude $\varphi_d$ and longitude $\lambda_d$, the local east, north, and
up directions may be written as

\begin{subequations}
\label{eq:local-basis-supp}
\begin{align}
 \hat{\bm e}_{E}
 &=(-\sin\lambda_d,\;\cos\lambda_d,\;0),\\
 \hat{\bm e}_{N}
 &=(-\sin\varphi_d\cos\lambda_d,
   -\sin\varphi_d\sin\lambda_d,\;\cos\varphi_d),\\
 \hat{\bm e}_{U}
 &=(\cos\varphi_d\cos\lambda_d,
   \cos\varphi_d\sin\lambda_d,\;\sin\varphi_d).
\end{align}
\end{subequations}
If the two interferometer arms have azimuths $\gamma_x$ and $\gamma_y$,
measured east of north, their unit vectors in this Earth-fixed frame are
\begin{subequations}
\label{eq:arm-vectors-supp}
\begin{align}
 \hat{\bm u}
 &=\cos\gamma_x\,\hat{\bm e}_{N}
   +\sin\gamma_x\,\hat{\bm e}_{E},\\
 \hat{\bm v}
 &=\cos\gamma_y\,\hat{\bm e}_{N}
   +\sin\gamma_y\,\hat{\bm e}_{E}.
\end{align}
\end{subequations}
For an ideal right-angle interferometer,
$\gamma_y-\gamma_x=\zeta=\pi/2$.  The corresponding detector tensor is
\begin{equation}
 D^{ij}_0=\frac{1}{2}
 \left(u^i u^j-v^i v^j\right).
 \label{eq:detector-tensor-supp}
\end{equation}
Thus the detector latitude, longitude, opening angle, and the directions of
its two arms are all contained in $D^{ij}_0$.  In the numerical calculation
we use the registered H1 location and arm geometry returned by PyCBC's
\texttt{Detector("H1")} object, with the H1 interferometer described in
Ref.~\cite{Aasi:2014mqd}; none of these geometric quantities is fitted from
the signal.

In a celestial frame the detector rotates rigidly about
$\hat{\bm Z}$.  Writing $\Phi(t)\equiv{\rm GMST}(t)$,
\begin{equation}
 D^{ij}(t)=R^i{}_{k}[\Phi(t)]R^j{}_{l}[\Phi(t)]D^{kl}_0,
 \label{eq:rotating-detector-tensor-supp}
\end{equation}
where $R(\Phi)$ is a rotation about the Earth's spin axis.  Each element of
$R$ is linear in $\{1,\cos\Phi,\sin\Phi\}$, so the rank-2 tensor
$D^{ij}(t)$ contains only the constant, first, and second sidereal harmonics.
With the ordering used in the Letter we may therefore write
\begin{equation}
 D^{ij}(t)=\sum_{n=1}^{5}{\cal D}^{ij}_n e_n(t),
 \label{eq:detector-harmonic-tensors-supp}
\end{equation}
with
\begin{equation}
 e_n(t)\in
 \left\{
 \cos 2\Phi(t),\;\sin 2\Phi(t),\;
 \cos \Phi(t),\;\sin \Phi(t),\;1
 \right\}.
 \label{eq:jks-basis-supp}
\end{equation}
The five constant tensors ${\cal D}^{ij}_n$ depend only on the detector
geometry in Eqs.~\eqref{eq:local-basis-supp}--\eqref{eq:detector-tensor-supp}.
This is the geometric reason that the expansion terminates after five terms.

To see how the sky position enters, let the source direction be
\begin{equation}
 \hat{\bm n}
 =\left(\cos\delta\cos\alpha,
         \cos\delta\sin\alpha,
         \sin\delta\right),
\end{equation}
and choose the $\psi=0$ basis on the plane of the sky as
\begin{subequations}
\begin{align}
 \hat{\bm p}&=(-\sin\alpha,\;\cos\alpha,\;0),\\
 \hat{\bm q}&=(-\sin\delta\cos\alpha,
                -\sin\delta\sin\alpha,\;\cos\delta).
\end{align}
\end{subequations}
The corresponding polarization tensors are
\begin{equation}
 e^+_{ij}=p_i p_j-q_i q_j,
 \qquad
 e^\times_{ij}=p_i q_j+q_i p_j.
 \label{eq:pol-tensors-supp}
\end{equation}
At $\psi=0$ the JKS functions are the contractions of the rotating detector
tensor with these two fixed sky tensors.  
Equivalently,
\begin{subequations}
\label{eq:AnBn-geometric-supp}
\begin{align}
 A_n(\alpha,\delta)
    &=\frac{{\cal D}^{ij}_n e^+_{ij}(\alpha,\delta)}{\sin\zeta},\\
 B_n(\alpha,\delta)
    &=\frac{{\cal D}^{ij}_n e^\times_{ij}(\alpha,\delta)}{\sin\zeta}.
\end{align}
\end{subequations}
Hence $A_n$ and $B_n$ contain two pieces and only two pieces: the fixed site
and arm geometry through ${\cal D}^{ij}_n$, and the source direction through
$e^+_{ij}$ and $e^\times_{ij}$.  They contain no waveform parameters, no
inclination, no luminosity distance, and no polarization angle.  
In
particular,
\begin{subequations}
 \label{eq:ab-expand}
\begin{align}
 a(t)&=\sum_{n=1}^{5}A_n e_n(t), \\
 b(t)&=\sum_{n=1}^{5}B_n e_n(t).
\end{align}
\end{subequations}
For a general polarization angle $\psi$, the antenna patterns are
\begin{subequations}
\label{eq:jks-pol}
\begin{align}
 F_+(t) &= \sin\zeta\,[a(t)\cos 2\psi+b(t)\sin 2\psi],\\
 F_\times(t) &= \sin\zeta\,[b(t)\cos 2\psi-a(t)\sin 2\psi].
\end{align}
\end{subequations}
For the H1 detector, $\zeta=\pi/2$ and the prefactor is unity.

\begin{figure}[t]
\centering
\includegraphics[width=0.49\textwidth]{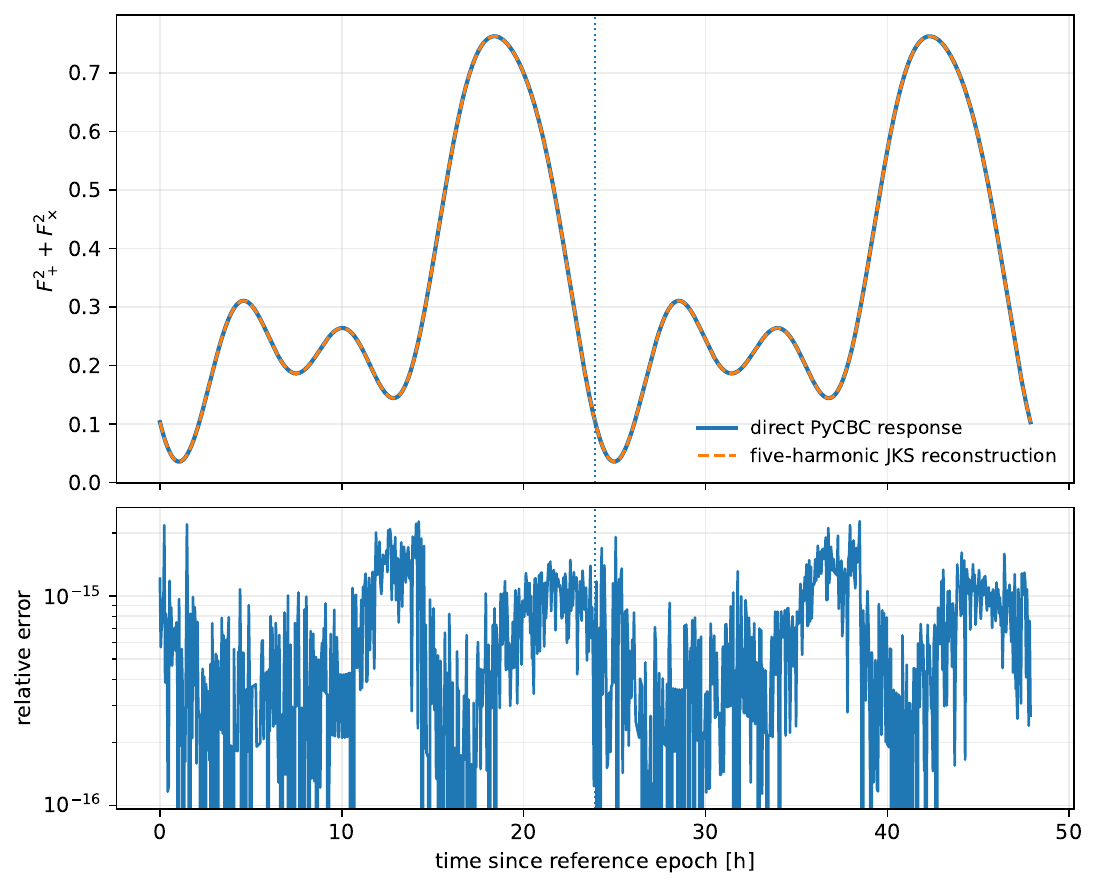}
\caption{Direct PyCBC antenna power $F_+^2+F_\times^2$ and its five-harmonic
JKS reconstruction for the LIGO-Hanford (H1) detector.  
The coefficients $A_n$ and $B_n$ contain the H1
site and arm geometry together with the source direction, as made explicit in
Eq.~\eqref{eq:AnBn-geometric-supp}.  They are fit on the first sidereal day,
to the left of the vertical dotted line; the second day uses the same
coefficients without refitting.  The lower panel shows the relative error.
The maximum absolute and relative errors are $1.11\times10^{-15}$ and
$2.28\times10^{-15}$, respectively.}
\label{fig:jks-check}
\end{figure}

The right-ascension dependence can be separated further.  In the
implementation we determine the coefficients at a fixed reference right
ascension $\alpha_{\rm ref}$ and obtain any other right ascension by rotating
the harmonic pairs.  If $\Delta\alpha=\alpha-\alpha_{\rm ref}$ and
\begin{equation}
 {\bf R}(\beta)=
 \begin{pmatrix}
  \cos\beta & -\sin\beta\\
  \sin\beta &  \cos\beta
 \end{pmatrix},
\end{equation}
then
\begin{subequations}
\begin{align}
 \binom{A_1}{A_2}_{\!\alpha}
    &= {\bf R}(2\Delta\alpha)
        \binom{A_1}{A_2}_{\!\alpha_{\rm ref}}, \\
 \binom{A_3}{A_4}_{\!\alpha}
    &= {\bf R}(\Delta\alpha)
        \binom{A_3}{A_4}_{\!\alpha_{\rm ref}}, \\
 A_5(\alpha) &= A_5(\alpha_{\rm ref}),
\end{align}
\end{subequations}
with identical relations for the $B_n$.  Thus, at fixed declination, changing
right ascension does not require a new harmonic fit.  It only changes the
phase of the $m=1$ and $m=2$ coefficient pairs.

For the nonprecessing, dominant-$(2,2)$ signals considered in the Letter, the
two polarizations are proportional to a common intrinsic waveform.  Using
\begin{align}
 C_+ &= \frac{1+\cos^2\iota}{2D_L},
 \\
 C_\times &= -\frac{i\cos\iota}{D_L},
\end{align}
the complex antenna amplitude is
\begin{align}
 {\cal F}(t)
    &= F_+(t)C_+ + F_\times(t)C_\times \nonumber \\
    &= \sin\zeta\sum_{n=1}^{5}G_n e_n(t),
 \label{eq:complex-jks}
\end{align}
where
\begin{equation}
 G_n=
 (A_n C_+ + B_n C_\times)\cos 2\psi
 +(B_n C_+ - A_n C_\times)\sin 2\psi.
 \label{eq:Gn-explicit}
\end{equation}
All time dependence is therefore carried by the five known functions $e_n$.
The detector geometry and sky direction are carried by $A_n,B_n$, while
$\psi$, $\iota$, and $D_L$ enter only through the inexpensive coefficients
$G_n$.

In practice we obtain $A_n$ and $B_n$ directly from the reference detector
implementation rather than transcribing closed-form coefficient expressions.
Setting $\psi=0$ gives $F_+=\sin\zeta\,a$ and
$F_\times=\sin\zeta\,b$.  At a fixed declination and reference right
ascension we sample the direct PyCBC antenna response at 50 times spanning one
sidereal day, form the $50\times5$ design matrix
$E_{kn}=e_n(t_k)$, and solve
\begin{equation}
 E\bm A=\frac{\bm F_+}{\sin\zeta},
 \qquad
 E\bm B=\frac{\bm F_\times}{\sin\zeta}
 \label{eq:fit-AnBn-supp}
\end{equation}
by linear least squares.  This is a tenfold-overdetermined fit for each
five-component vector.  The fit is only a numerical way of evaluating the
geometric contractions in Eq.~\eqref{eq:AnBn-geometric-supp}; it does not add
new degrees of freedom to the likelihood.

Figure~\ref{fig:jks-check} uses H1 geometry and a held-out sky position
$(\alpha,\delta)=(2.1,-0.3)$ with $\psi=0.77$.  The coefficients are fit using
the first sidereal day only and then held fixed over the second day.  The
maximum absolute and relative errors in $F_+^2+F_\times^2$ are
$1.11\times10^{-15}$ and $2.28\times10^{-15}$, respectively.

\section{Fixed stationary-time map and accelerated likelihood}
\label{sec:likelihood-construction}

This section gives the implementation details behind
\emph{The stationary-time map}, \emph{Intrinsic/extrinsic factorization},
and \emph{Curvature-controlled grid and acceleration by relative binning}
in the Letter, together with the numerical values underlying Fig.~1.
The stationary-time prescription uses the standard stationary-phase
treatment~\cite{Sathyaprakash:1991mt,Cutler:1994ys}.  Related
relative-binning constructions for more general waveforms and long-duration
3G signals are discussed in
Refs.~\cite{Narola2024,KumarGuptaSathyaprakash2025,Baral2025}.

The sidereal basis and the time-dependent detector delay are evaluated along a
stationary-time map constructed once from the full waveform phase at the
fiducial intrinsic point $\bm\lambda_0$.  During sampling we make the
replacement
\begin{equation}
 t_f(f;\bm\lambda)\longrightarrow t_f^{\bm\lambda_0}(f)
 \label{eq:fixed-map-replacement}
\end{equation}
inside these slowly varying response terms.  The intrinsic waveform
$\tilde h_{22}(f;\bm\lambda)$ is still evaluated at every trial intrinsic
point.  A Newtonian estimate over a representative MM$=0.95$ local domain
gives a maximum sidereal-phase displacement of $4.3\times10^{-5}$ rad at
$\flow=2$ Hz.  The direct likelihood test below is substantially more
restrictive.

The production calculation uses H1 geometry~\cite{Aasi:2014mqd} with
the ET-D PSD~\cite{Hild2011} over $[2,1800]$ Hz.  The curvature-grid tolerance is
$\epsilon_{\rm grid}=0.1$ and the relative-binning tolerance is
$\epsilon_b=0.05$.  At $\flow=2$ Hz the curvature grid contains
$8{,}379{,}287$ in-band frequency samples and the online likelihood uses
$10{,}295$ relative bins.  The fixed-map tolerance used to construct the
model-phase stationary-time interpolant is $0.05$ s.

For a trial intrinsic point we define
\begin{equation}
 \rho(f;\bm\lambda)
 =\frac{\tilde h_{22}(f;\bm\lambda)}
 {\tilde h_{22}(f;\bm\lambda_0)},
 \qquad
 r(f)=\rho(f;\bm\lambda)e^{-2\pi i f\Delta\tau(f)},
 \label{eq:rb-ratios-supp}
\end{equation}
with
\begin{equation}
 \Delta\tau(f)=
 \tau+\Delta t\!\left[\alpha,\delta;
 t_f^{\bm\lambda_0}(f)\right].
\end{equation}
The JKS split removes the sky-dependent antenna amplitude from the expensive
frequency sums.  The remaining coalescence-time and Doppler phases are applied
at the sparse relative-bin edges.

The local accuracy benchmark in Fig.~1 of the Letter uses 31 nearby trial
points together with the injection.  At $\flow=2$ Hz and
$\epsilon_b=0.05$, the centered absolute likelihood error has median
$0.24494$, 90th percentile $0.33417$, and maximum $0.38468$.  Tightening the
binning relative to $\epsilon_b=0.085$ reduces the corresponding median from
$0.41640$ to $0.24494$.  The 2-Hz median speed-up for the production tolerance
is $1.5003\times10^4$, with an interquartile range
$[1.4997,1.5025]\times10^4$.

\section{Posterior-support native-likelihood validation}
\label{sec:posterior-native}

This section gives the native-likelihood check referred to in
\emph{Demonstration} in the Letter, following the pooled-posterior result.
It tests the accelerated likelihood over the actual posterior support and
separates the errors due to the fixed stationary-time map, curvature grid,
and relative-binning approximation.

The local benchmark points are deliberately broader than the posterior.  We
therefore perform a second validation directly over posterior support.  The
50 independent sampling runs are first combined with equal total weight per
run.  From the resulting pooled posterior we select 1000 samples, balanced
across the 50 sampler seeds, and evaluate four likelihoods at each point:

\begin{table}[h!]
\centering
\begin{tabular}{ll}
\hline\hline
Quantity & Definition \\
\hline
$\ln\Like_{\rm RB}$ &
Production JKS+relative-binning likelihood \\
$\ln\Like_{\rm curv}$ &
Direct likelihood on the curvature grid \\
$\ln\Like_{\rm fixed}$ &
Native grid with the fixed fiducial map \\
$\ln\Like_{\rm native}$ &
Native grid with a trial-specific map \\
\hline\hline
\end{tabular}
\caption{Likelihood variants used in the validation.}
\label{tab:likelihood-variants-supp}
\end{table}

The native calculations are streamed in blocks of $2^{20}$ frequency samples,
so the duration-resolution grid is never stored as one full array.

For any contribution $X$, the quoted error is centered at the injection,
\begin{equation}
 \delta\ln\Like_X(\bm\theta)
 =\Delta\ln\Like_X(\bm\theta)
 -\Delta\ln\Like_X(\bm\theta_{\rm inj}).
 \label{eq:centered-error-supp}
\end{equation}
The error decomposition is
\begin{subequations}
\label{eq:error-decomposition}
\begin{align}
 \Delta\ln\Like_{\rm total}
 &=\ln\Like_{\rm RB}-\ln\Like_{\rm native},\\
 \Delta\ln\Like_{\rm map}
 &=\ln\Like_{\rm fixed}-\ln\Like_{\rm native},\\
 \Delta\ln\Like_{\rm curv}
 &=\ln\Like_{\rm curv}-\ln\Like_{\rm fixed},\\
 \Delta\ln\Like_{\rm relbin}
 &=\ln\Like_{\rm RB}-\ln\Like_{\rm curv}.
\end{align}
\end{subequations}
The four terms therefore separate the fixed-map, curvature-grid, and
relative-binning approximations while retaining the full production error as
their sum.

\begin{table*}[htbp]
\label{tab:error-budget}
\centering
\caption{Posterior-support centered likelihood errors from 1000 samples.}
\begin{tabular}{lrrrr}
\toprule
Contribution & Median & $P_{90}$ & $P_{95}$ & Maximum \\
\midrule
Total
& $3.46\!\times\!10^{-3}$
& $8.13\!\times\!10^{-3}$
& $9.73\!\times\!10^{-3}$
& $1.79\!\times\!10^{-2}$ \\
Fixed map
& $6.13\!\times\!10^{-6}$
& $3.38\!\times\!10^{-5}$
& $4.92\!\times\!10^{-5}$
& $1.70\!\times\!10^{-4}$ \\
Curvature grid
& $4.08\!\times\!10^{-5}$
& $2.13\!\times\!10^{-4}$
& $2.89\!\times\!10^{-4}$
& $7.79\!\times\!10^{-4}$ \\
Relative binning
& $3.48\!\times\!10^{-3}$
& $8.13\!\times\!10^{-3}$
& $9.77\!\times\!10^{-3}$
& $1.80\!\times\!10^{-2}$ \\
\bottomrule
\end{tabular}
\end{table*}

\begin{figure*}[t]
\centering
\includegraphics[width=0.9\textwidth]{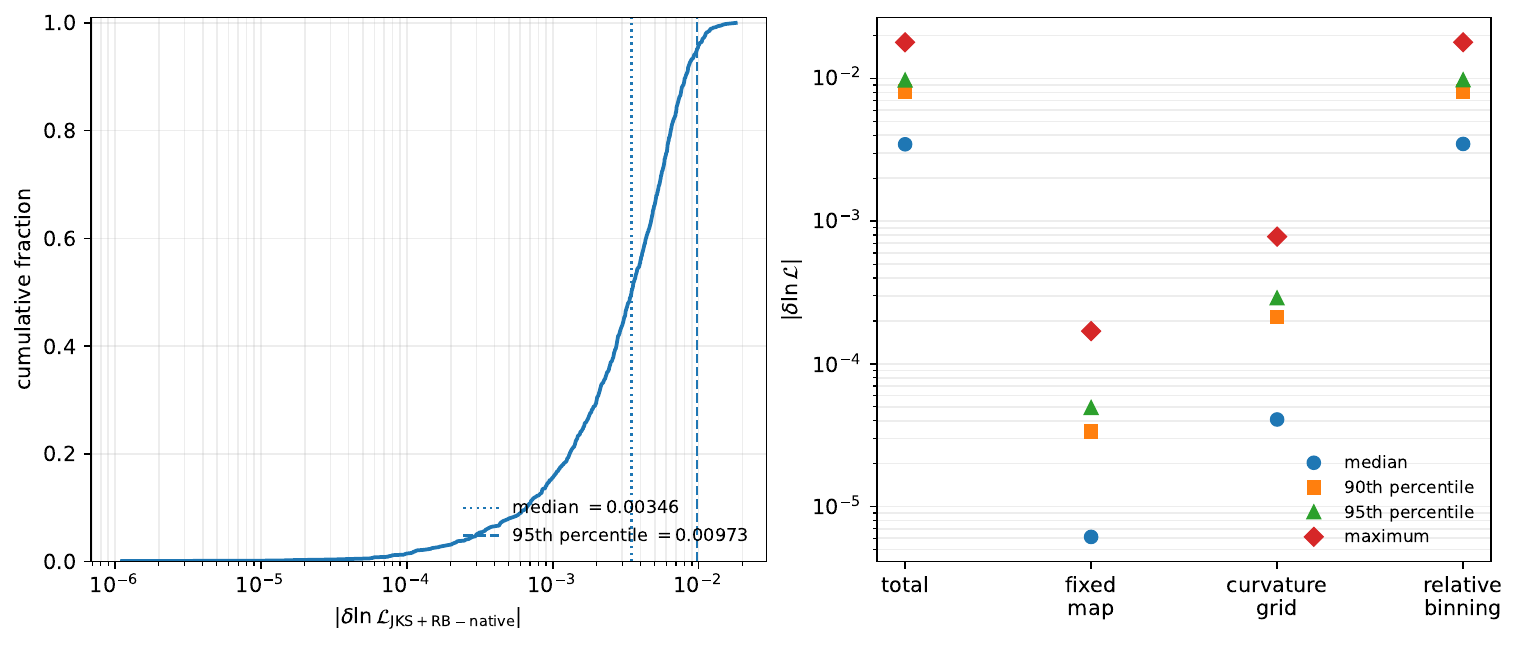}
\caption{Posterior-support likelihood validation using 1000 samples from the
equal-run-weight pooled posterior.  Left: empirical cumulative distribution
of the centered difference between the production JKS+relative-binning
likelihood and the streamed native-resolution likelihood with a
trial-specific stationary-time map.  Right: median, 90th percentile, 95th
percentile, and maximum absolute centered errors for the full likelihood and
for the fixed-map, curvature-grid, and relative-binning contributions defined
in Eq.~\eqref{eq:error-decomposition}.}
\label{fig:posterior-validation}
\end{figure*}

The full accelerated-native error has median
$3.461\times10^{-3}$, 95th percentile $9.726\times10^{-3}$, and maximum
$1.794\times10^{-2}$.  The fixed-map contribution is smaller by about three
orders of magnitude in the median, with maximum $1.70\times10^{-4}$.  The
curvature-grid contribution is also below $10^{-3}$ at every sampled point.
The residual error is therefore set almost entirely by relative binning.
Fig.~(\ref{fig:posterior-validation}) shows the full distribution and the
same error budget graphically.

The likelihood comparison above tests the accelerated likelihood point by
point over the posterior support.  A separate question is whether the
remaining likelihood error is large enough to change the inferred posterior.
We test this without performing a second native sampling run.  Since the
native likelihood has already been evaluated at the same 1000 posterior
samples, each sample can be assigned the importance weight
\begin{equation}
 w_i^{\rm native}
 \propto
 \exp\!\left[
 \ln\Like_{\rm native}(\bm\theta_i)
 -\ln\Like_{\rm RB}(\bm\theta_i)
 \right].
 \label{eq:native-reweight}
\end{equation}
If the accelerated and native likelihoods assign different relative
probabilities across the posterior, these weights will vary from sample to
sample and the reweighted parameter distributions will shift.  If the
likelihood correction is nearly constant over the posterior, the weights
remain nearly uniform.

For the 1000 samples used here, the effective sample size of the native
weights is
\begin{equation}
 N_{\rm eff}
 =
 \frac{1}{\sum_i (w_i^{\rm native})^2}
 =
 999.976,
 \label{eq:rewt-ess}
\end{equation}
out of 1000 samples, corresponding to an ESS fraction of $0.999976$.
Thus the native likelihood correction changes the relative weights of these
posterior samples only very weakly.

Table~\ref{tab:reweight} compares the same 1000 samples before and after
applying the native-likelihood weights.  Because the sample locations are
identical in the two columns, their difference isolates the effect of replacing
$\Like_{\rm RB}$ by $\Like_{\rm native}$.  The median shifts are below
$1.5\times10^{-3}$ of the 16th--84th percentile half-width of the full pooled
posterior for all four quantities shown.

\begin{table*}[t]
\caption{Comparison of the same 1000 posterior samples before and after
native-likelihood reweighting. Since the sample locations are unchanged,
the difference between the two median columns is due only to the likelihood
correction in Eq.~\eqref{eq:native-reweight}.}
\label{tab:reweight}
\centering
\begin{tabular}{lccc}
\toprule
Quantity & Selected 1000 median & Native-reweighted median
& Shift / full 68\% half-width \\
\midrule
$\Mc^{\rm det}$ [$\Msun$]
& $1.18484448811$ & $1.18484448822$ & $+2.45\times10^{-4}$ \\
$q$
& $0.8699165007$ & $0.8699160456$ & $-1.17\times10^{-4}$ \\
$A_{\rm eff}$ [Mpc$^{-1}$]
& $2.815447\times10^{-3}$ & $2.815806\times10^{-3}$ & $+1.47\times10^{-3}$ \\
$\Delta t_c^{\rm geo}$ [ms]
& $0.591516$ & $0.589641$ & $-1.48\times10^{-3}$ \\
\bottomrule
\end{tabular}
\end{table*}
In Table~\ref{tab:reweight}, 
$
 A_{\rm eff}\equiv
 \sqrt{\left({1+\cos^2\iota}\right)^2/4+\cos^2\iota}\ / D_L
$
is the single-detector amplitude combination used in the sampler-stability
comparison. 

Thus, over the posterior support tested here, replacing the
accelerated likelihood with the native likelihood leaves the inferred
posterior essentially unchanged.

\begin{figure*}[tb]
\centering
\includegraphics[width=0.90\textwidth]{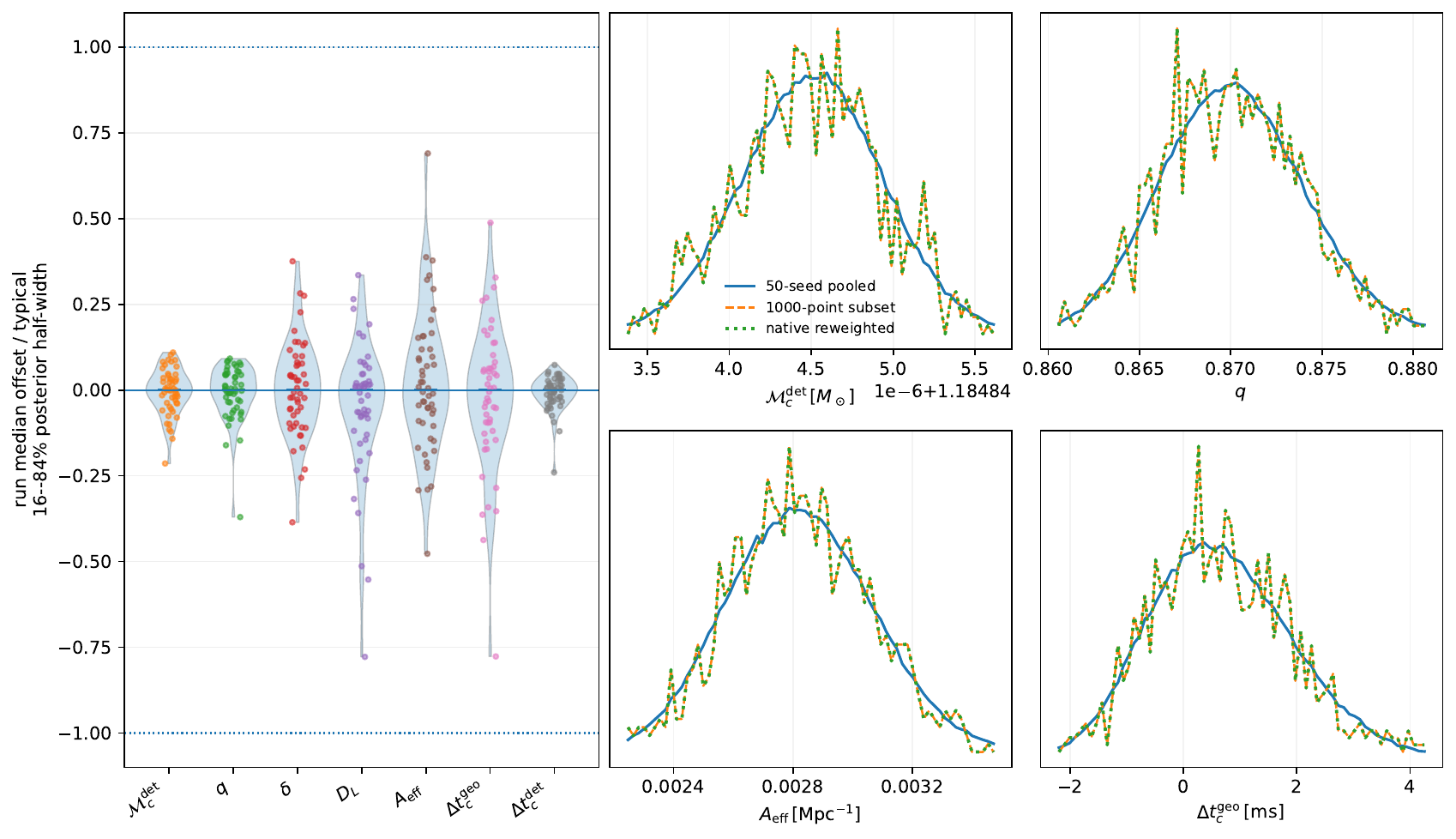}
\caption{Sampler-seed stability and native-likelihood reweighting.  Left:
median offsets for 50 independent \textsc{dynesty} runs, normalized by the
typical 16th--84th percentile half-width for each branch-stable quantity.  The
dotted lines mark one posterior half-width.  Right: the full equal-run-weight
pooled posterior, the 1000 samples used for the native check, and those same
1000 samples after native-likelihood reweighting.  The effective sample size
after reweighting is $999.976$ out of 1000.}
\label{fig:stability-reweight}
\end{figure*}

\section{Parameter-estimation configuration and sampler stability}
\label{sec:pe-details}

This section records the configuration underlying Fig.~2 of the Letter and
the timing and sampler-stability statements in \emph{Demonstration}.  The
parameter-estimation example uses zero-noise data, H1
geometry~\cite{Aasi:2014mqd}, the ET-D PSD~\cite{Hild2011}, and the
$[2,1800]$ Hz analysis band.  The sampled coordinates are
\begin{equation}
 (\Mc^{\rm det},q,\alpha,\delta,\psi,\iota,D_L,t_c),
\end{equation}
with the coalescence phase marginalized analytically.  The injection has
${\Mc^{\rm det}=1.1848446629484104\,\Msun}$,
${q=0.8698630136986302}$, ${(\alpha,\delta)=(1.2,0.4)}$,
$\psi=0.6$, $\iota=0.9$, $D_L=300$ Mpc, and
${t_c=1126259462.0}$ s.  Table~\ref{tab:priors} lists the priors.

\begin{table*}
\caption{Injection values and priors for the eight sampled parameters.}
\label{tab:priors}
\centering
\begin{tabular}{lccc}
\toprule
Parameter & Injection & Prior & Range \\
\midrule
$\Mc^{\rm det}$ [$\Msun$]
& $1.18484466295$ & uniform
& $\mathcal{M}_{c,{\rm inj}}^{\rm det}(1\pm2\times10^{-5})$ \\
$q$
& $0.869863014$ & uniform & $[0.789863014,\,0.949863014]$ \\
$\alpha$
& $1.2$ & uniform & $[0,2\pi)$ \\
$\delta$
& $0.4$ & isotropic sky & $[-\pi/2,\pi/2]$ \\
$\psi$
& $0.6$ & uniform & $[0,\pi/2)$ \\
$\iota$
& $0.9$ & isotropic orientation & $[0,\pi]$ \\
$D_L$ [Mpc]
& $300$ & uniform volume & $[30,3000]$ \\
$t_c$ [s]
& $1126259462.0$ & uniform & $t_{c,{\rm inj}}\pm0.01$ \\
\bottomrule
\end{tabular}
\end{table*}

Each production run uses \textsc{dynesty}~\cite{Speagle2020} with
2048 live points, \texttt{bound=multi}, and random slice sampling
(\texttt{sample=rslice})~\cite{Neal2003}, with \texttt{bootstrap=0}, and
stopping criterion $\Delta\ln Z=0.1$.  Right ascension and polarization are
treated as periodic parameters.  The production configuration uses 48 worker
processes with queue size 48.  Fifty independent seeds, 7908--7957, are run
with otherwise identical settings.  Their equal-run-weight pooled posterior
contains 200,000 samples.

Across the 50 runs, the median sampling time is \mbox{$311.6$ s}, with 16th--84th
percentiles $305.2$--$318.1$ s.  The median number of likelihood calls is
$3.582\times10^6$, and the median weighted effective sample size is
$1.386\times10^4$.  The median one-time likelihood build time is $11.16$ s.
For the branch-stable quantities used in Fig.~\ref{fig:stability-reweight},
the largest displacement of an individual-run median from the ensemble median
is $0.778$ times the typical 16th--84th percentile half-width; this occurs for
$D_L$.  Raw $\iota$ and $\psi$ are excluded from this single summary statistic
because their medians are branch-sensitive in the single-detector orientation
degeneracy.

\end{document}